\documentclass[journal,comsoc]{IEEEtran}
\usepackage[T1]{fontenc}% optional T1 font encoding

\ifCLASSINFOpdf
\else
\fi
\usepackage{amsmath}
\usepackage[cmintegrals]{newtxmath}
\usepackage{graphicx}
\usepackage{hyperref}
\usepackage{algorithm}
\usepackage{algpseudocode}
\usepackage[noadjust]{cite}
\DeclareMathOperator*{\argmin}{argmin}

\newcommand{\etal}{\textit{et al}}
\newcommand{\Fig}{\text{Fig. }}

\usepackage{color}

\newcommand*{\blue}{\textcolor{blue}}
\begin{document}
\title{Multipass SAR Interferometry Based on Total Variation Regularized Robust Low Rank Tensor Decomposition}

\author{Jian~Kang,~\IEEEmembership{Member,~IEEE,}
        Yuanyuan~Wang,~\IEEEmembership{Member,~IEEE,}
        and~Xiao~Xiang~Zhu~\IEEEmembership{Senior~Member,~IEEE}% <-this % stops a space
\thanks{This work is jointly supported by the European Research Council (ERC) under the European Union's Horizon 2020 research and innovation program (ERC-2016-StG-714087, So2Sat) and by Helmholtz Association under the framework of the Young Investigators Group ``SiPEO'' (VH-NG-1018, www.sipeo.bgu.tum.de), Helmholtz Artificial Intelligence Cooperation Unit (HAICU) - Local Unit ``Munich Unit @Aeronautics, Space and Transport (MASTr)'' and Helmholtz Excellent Professorship ``Data Science in Earth Observation - Big Data Fusion for Urban Research''.}
\thanks{\emph{Corresponding author: Xiao Xiang Zhu}}
\thanks{J. Kang is with Signal Processing in Earth Observation (SiPEO), Technical University of Munich (TUM), 80333 Munich, Germany (e-mail: jian.kang@tum.de)}% <-this % stops a space
\thanks{Y. Wang and X. X. Zhu  are with the Remote Sensing Technology Institute (IMF), German Aerospace Center (DLR), 82234 Wessling, Germany, and also with Signal Processing in Earth Observation (SiPEO), Technical University of Munich (TUM), 80333 Munich, Germany (e-mail: y.wang@tum.de; xiaoxiang.zhu@dlr.de)}}% <-this % stops a space


\markboth{IEEE TRANSACTIONS ON GEOSCIENCE AND REMOTE SENSING}%
{\MakeLowercase{\textit{Kang et al.}}: Multipass SAR Interferometry...}



\maketitle

\begin{abstract}
\blue{This is the pre-acceptance version, to read the final version,  please  go  to  IEEE  Transactions  on  Geoscience  and Remote Sensing on IEEE Xplore.} Multipass SAR interferometry (InSAR) techniques based on meter-resolution spaceborne SAR satellites, such as TerraSAR-X or COSMO-Skymed, provide 3D reconstruction and the measurement of ground displacement over large urban areas. Conventional method such as Persistent Scatterer Interferometry (PSI) usually requires a fairly large SAR image stack (usually in the order of tens), in order to achieve reliable estimates of these parameters. Recently, low rank property in multipass InSAR data stack was explored and investigated in our previous work \cite{kang2018romio}. By exploiting this low rank prior, more accurate estimation of the geophysical parameters can be achieved, which in turn can effectively reduce the number of interferograms required for a reliable estimation. Based on that, this paper proposes a novel tensor decomposition method in complex domain, which jointly exploits low rank and variational prior of the interferometric phase in InSAR data stacks. Specifically, a total variation (TV) regularized robust low rank tensor decomposition method is exploited for recovering outlier-free InSAR stacks. We demonstrate that the filtered InSAR data stacks can greatly improve the accuracy of geophysical parameters estimated from real data. Moreover, this paper demonstrates for the first time in the community that tensor-decomposition-based methods can be beneficial for large-scale urban mapping problems using multipass InSAR. Two TerraSAR-X data stacks with large spatial areas demonstrate the promising performance of the proposed method.      
\end{abstract}


% \fi
\IEEEpeerreviewmaketitle

\section{Introduction}

\subsection{Multipass InSAR}

With respect to different scattering cases, i.e. point scatterers and distributed scatterers, methods for the retrieval of geophysical parameters (namely elevation and deformation parameters) for large areas can be accordingly split into two categories: Persistent Scatterer Interferometry (PSI) \cite{ferretti2001permanent,adam2003development,fornaro2009deformation,sousa2011persistent,gernhardt2012deformation,kampes2006radar,wang2014efficient,costantini2014persistent,zhang2011modeling,de2009detection} and Distributed Scatterer Interferometry (DSI) \cite{ferretti2011new,goel2012advanced,Wang201289,jiang2015fast,samiei2016phase,wang2016robust,cao2016phase}. Those methods are the backbone of data analysis based on multipass InSAR stacks and widely exploited for 3D urban reconstruction and surface displacement monitoring.

Generally, the key steps of PSI \cite{ferretti2001permanent,adam2003development,fornaro2009deformation,sousa2011persistent,gernhardt2012deformation,kampes2006radar,wang2014efficient,costantini2014persistent,zhang2011modeling,de2009detection,crosetto2015measuring,devanthery2014approach} involve PS candidate identification and parameter estimation. For example, PS pixels can be selected according to \textit{amplitude dispersion index}, which can be calculated by the ratio between the temporal standard deviation and mean of the amplitudes \cite{ferretti2001permanent}.  By exploiting the spatial correlation of phase measurements, Stanford method for persistent scatterers (StaMPS) \cite{hooper2005persistent} is applicable for selecting PS in areas undergoing non-steady deformation without prior knowledge. Likewise, based on spatial correlation analysis, PS pairs are identified via the construction of PS arc in \cite{costantini2008new}. Sublook coherence approached is proposed in \cite{schneider2006polarimetric} for point-like scatterer identification without the requirement of certain number of temporal SAR images. Methods for estimating geophysical parameters such as topography height and linear deformation rates from PS are usually based on maximum likelihood estimator (MLE) \cite{ferretti2001permanent}. In order to describe the precision of the estimated parameters, least squares ambiguity decorrelation (LAMBDA), which is originally developed for the ambiguity resolution of GPS signal, is adapted to parameter estimation for PS signals in \cite{kampes2004ambiguity}. When layover phenomenon is taken into account, differential SAR tomography (D-TomoSAR) \cite{Fornaro2003,lombardini2005,zhu2010very,zhunolinear2011,reale2011,fornaro2014tomographic,budillon2015glrt} was proposed for efficiently reconstructing the real 3D structure of the scene. Such technique mainly contains two steps: identification of pixels with multiple PSs and parameter estimation based on tomographic inversion.

In order to extract geophysical information from non-urban areas with DS, interferometry techniques for parameter estimation from such stochastic signals have been extensively carried out since a decade ago. Normally, statistically homogeneous pixel (SHP) selection for covariance matrix estimation and optimal phase history retrieval from such covariance matrices are the two key steps in DS interferometry. As introduced in \cite{ferretti2011new}, SqueeSAR exploits  Kolmogorov-Smirnov (KS) test for selecting SHP with the assumption that the statistics of amplitude data can be seen as a proxy for phase stability. Composed of KS, Anderson-Darling (AD), Kullback-Leibler divergence and generalized likelihood-ratio test (GLRT), different amplitude-based methods for selecting SHP are evaluated in \cite{parizzi2011adaptive}. Estimating optimal phase histories from covariance matrices built by the selected SHP is the second key step in DSI. The construction of covariance matrices can be considered as the generation of multimaster (MM) interferograms. In order to link all the available interferometric phases, optimal phase histories, i.e. SM phases, are then estimated from such covariance matrices. It is also well-known as \textit{phase linking} or \textit{phase triangulation} \cite{guarnieri2008exploitation, ferretti2011new,samiei2016phase,wang2016robust}. Then, the corresponding geophysical parameters can be reconstructed in a similar processing chain of PS signals. 

Although those conventional techniques for geophysical parameter estimation do exploit information from multiple neighbouring pixels, no explicit semantic and geometric information that might be preserved in the images has been utilized. Recently, several multipass InSAR techniques have been developed based on exploiting semantic and geometric information preserved in SAR images for improving geophysical parameter estimation. In \cite{zhumslimmer}, Zhu \etal. demonstrated that by introducing building footprints from OpenStreetMap (OSM) as prior knowledge of pixels sharing similar heights into frameworks based on joint sparse reconstruction techniques, a highly accurate tomographic reconstruction can be achieved using just six interferograms, instead of the typically-required 20-100. \cite{ferraioli2018parisar} proposed a method for multi-baseline InSAR phase unwrapping based on combining non-local denoising methods and total variation regularized spectral estimation method. In our previous work, a general framework for object-based InSAR deformation reconstruction based on a tensor-model with a regularization term is proposed. It makes use of external semantic labels of various objects like bridges, roofs and fa\c{c}ades, as an input for the support of the total variation regularizer \cite{kang2017robust,kang2016object}. However, it requires explicit and fairly accurate semantic labels for a reliable performance. Therefore, \cite{kang2018romio} investigated the inherent low rank property of multipass InSAR phase tensors. It allows loose semantic labels, such as a rectangle covering major part of an object, for object-based geophysical parameter reconstruction in urban areas.

As a follow-on work, we seek to develop a novel method for parameter retrieval from multipass InSAR data stacks by jointly considering the variational prior \cite{kang2017robust} and the low rank property \cite{kang2018romio} of InSAR stacks. To this end, a total variation (TV) regularized robust low rank tensor decomposition method in complex domain is proposed in this paper, in order to recover outlier-free InSAR data stacks.

\begin{table*}
	\caption{Mathematic notations}
	\centering
	\renewcommand{\arraystretch}{1.3}
	\begin{tabular}{l | l}
		\hline
		$ \mathcal{X}, \mathbf{X}, \mathbf{x}, x $				& tensor, matrix, vector, scalar \\
		$ \mathbf{X}_{(n)} $									& mode-$ n $ unfolding of tensor $ \mathcal{X} $ \\
		$ \langle\mathcal{X},\mathcal{Y}\rangle $				& inner product of tensor $ \mathcal{X} $ and $ \mathcal{Y} $, i.e. the sum of product of their entries \\
		$ \|\mathcal{X}\|_F $									& Frobenius norm of tensor $ \mathcal{X} $, i.e. $ \|\mathcal{X}\|_F=\sqrt{\langle\mathcal{X},\mathcal{X}\rangle} $ \\
		$ \mathrm{vec}(\mathcal{X}) $							& vectorization of $\mathcal{X}$ \\
		$ \|\mathcal{X}\|_1 $									& $ L_1 $ norm of tensor $ \mathcal{X} $, i.e. $ \|\mathcal{X}\|_1=\|\mathrm{vec}(\mathcal{X})\|_1 $ \\
		$ \|\mathbf{X}\|_* $									& matrix nuclear norm: the sum of its singular values, i.e. \(\|\mathbf{X}\|_*:=\sum_{i}\sigma_{i}\) \\
		$ \otimes $                                             & outer product \\
		$ \odot $                                               & element-wise product \\
		\hline
	\end{tabular}
	\label{table:Mathematic_notation}
\end{table*}

\subsection{Contributions}
The contributions of this paper are summarized as follows:
\begin{itemize}
	\item Based on the prior knowledge of low rank and smoothness of multipass InSAR data stacks, a novel tensor decomposition method in complex domain is proposed, i.e. a total variation (TV) regularized robust low rank tensor decomposition, for recovering outlier-free InSAR data stacks.   
	\item The proposed method not only takes advantages of both variational prior \cite{kang2017robust} and the low rank property \cite{kang2018romio} of InSAR stacks, but also it can avoid the requirement of explicit semantic labels for object-based geophysical parameter reconstruction.
	\item This paper firstly presents tensor-decomposition-based methods can be beneficial for large-scale urban mapping problems, including 3D reconstruction and surface displacement monitoring.  
\end{itemize}
\subsection{Structure of this paper}
The rest of this paper is organized as follows. Section \ref{sc:recap} introduces the notations utilized in this paper and recaps our previous work. In Section \ref{sc:3DTVLR_method}, the proposed total variation regularized robust low rank tensor decomposition in complex domain is introduced, together with its optimization procedure. Simulated experiments are conducted in Section \ref{sc:simulation}. Case studies of large-scale real data in Berlin and Las Vegas are performed in Section \ref{sc:real_data}. Section \ref{sc:conclustion} draws the conclusion of this paper.

\section{Background knowledge}\label{sc:recap}

\subsection{Notations and Tensor Model of Multipass InSAR Data Stacks}

A tensor can be considered as a multi-dimensional array. The \textit{order} of a tensor is the number of its \textit{modes} or \textit{dimensions}. A tensor of order $ N $ in the complex domain can be denoted as $ \mathcal{X} \in \mathbb{C}^{I_1 \times I_2 \cdots \times I_N}$ and its entries as $ x_{i_1,i_2,\cdots,i_N} $. Specifically, vector $ \mathbf{x} $ is a tensor of order one, and matrix $ \mathbf{X} $ can be represented as a tensor of order two. \textit{Fibers} are the higher-order analogy of matrix rows and columns, which are defined by fixing every index but one. \textit{Slices} of a tensor are obtained by fixing all but two indices. Matricization, also known as \textit{unfolding}, is the process of reordering the elements of a tensor into a matrix. Specifically, the mode-$ n $ unfolding of tensor $ \mathcal{X} $ is defined by $ \mathbf{X}_{(n)} $ that is obtained by arranging the mode-$ n $ fibers as the columns of the matrix. The utilized tensor notations are summarized in Table \ref{table:Mathematic_notation}.

As proposed in our previous work \cite{kang2017robust,kang2018romio}, an InSAR data stack can be represented by a 3-mode tensor: $ \mathcal{G}\in \mathbb{C}^{I_{1}\times{I_{2}}\times{I_{3}}} $, where $ I_{1} $ and $ I_{2} $ represent the spatial dimensions in range and azimuth, and $ I_3 $ denotes the number of SAR interferograms. The InSAR data tensor can be modeled by
\begin{equation}
\overline{\mathcal{G}}(\mathbf{S},\mathbf{P})=\mathcal{A}\odot\exp\{-j(\frac{4\pi}{\lambda{r}}\mathbf{S}\otimes\mathbf{b}+\frac{4\pi}{\lambda}\mathbf{P}\otimes\boldsymbol{\tau})\},
\label{eq:multipass_InSAR_tensor_model}
\end{equation}
where $ \overline{\mathcal{G}} $ is the modeled InSAR data tensor, $ \mathcal{A} $ denotes the modeled amplitude tensor, $ \mathbf{b}\in\mathbb{R}^{I_3} $ is the vector of the spatial baselines,  $ \boldsymbol{\tau}\in\mathbb{R}^{I_3} $ is a warped time variable \cite{zhunolinear2011}, e.g. $ \boldsymbol{\tau}=\mathbf{t} $ for a linear motion, and $ \boldsymbol{\tau}=\sin(2\pi(\mathbf{t}-t_0)) $ for a seasonal motion model with temporal baseline $ \mathbf{t} $ and time offset $ t_0 $. $ \mathbf{S}\in \mathbb{R}^{I_1\times I_2} $ and $ \mathbf{P}\in \mathbb{R}^{I_1\times I_2} $ are the unknown elevation and deformation maps to be estimated, respectively, $ \lambda $ is the wavelength of the radar signals and $ r $ denotes the range between radar and the observed area.

\subsection{Multipass InSAR with TV Regularizer}
By integrating smoothness prior knowledge of deformation map into the parameter retrieval, \cite{kang2017robust} introduces a joint reconstruction model of object-based deformation parameters by exploiting TV regularization. Correspondingly, the object-based model can be summarized as:
\begin{equation}
\{\hat{\mathbf{S}},\hat{\mathbf{P}}\}=\argmin_{\mathbf{S},\mathbf{P}}\frac{1}{2}\|\mathcal{W}\odot(\mathcal{G}-\overline{\mathcal{G}}(\mathbf{S},\mathbf{P}))\|_F^2+\eta{f(\mathbf{S},\mathbf{P})},
\label{eq:5}
\end{equation} 
where $ \mathcal{G} $ is the observed InSAR data stack, $ \mathcal{W} $ denotes a weighting tensor, $ \eta $ is the penalty parameter for balancing the two terms in \eqref{eq:5} and $ f(\mathbf{S},\mathbf{P}) $ denotes the penalty term which represents the spatial prior of $ \mathbf{S} $ and $ \mathbf{P} $. Specifically, smoothness prior, such as TV norm, can be considered for urban area reconstruction. %However, explicit and accurate semantic labels for object areas are demanded in this work to achieve reliable results. 
\subsection{Low Rank Tensor Decomposition in Multipass InSAR}
Moreover, seeking to magnify the power of object-based method for multipass InSAR, we investigate the low rank property inherent in InSAR data stacks \cite{kang2018romio}, according to the following knowledge:
\begin{itemize}
	\item It can be generally assumed that the elevation and deformation maps, $ \mathbf{S} $ and $ \mathbf{P} $, follow certain regular structure or homogeneous pattern, because of the regular man-made structures in urban areas.
	\item The observed SAR images of urban object areas are usually highly correlated along the temporal dimension.
\end{itemize} 
By exploiting the low rank property, object-based InSAR data stacks can be robustly recovered based on robust low rank tensor decomposition:
\begin{equation}
\{\hat{\mathcal{X}},\hat{\mathcal{E}}\}=\argmin_{\mathcal{X},\mathcal{E}}\|\mathcal{X}\|_*+\gamma\|\mathcal{E}\|_1,\quad s.t.~\mathcal{X}+\mathcal{E}=\mathcal{G},
\label{eq:HoRPCA_convex}
\end{equation}
where $ \hat{\mathcal{X}} $ and $\hat{\mathcal{E}} $ are the recovered outlier-free InSAR data tensor and the estimated outlier tensor, respectively. Based on this model, \cite{kang2018romio} demonstrates that reliable parameter estimation can be maintained, given loose semantic labels of objects. However, smoothness structures of multipass InSAR data stacks are not exploited in the model \ref{eq:HoRPCA_convex}. As introduced in \cite{kang2017robust,ferraioli2017parisar,baselice2014contextual}, geophysical paramter estimation can be enhanced by considering smoothness structures of elevation and deformation maps.

\section{Combining Total Variation Regularized Robust Low Rank Tensor Decomposition} \label{sc:3DTVLR_method}

\subsection{Total Variation Regularized Robust Low Rank Tensor Decomposition}
To this end, we develop a novel tensor decomposition method in complex domain, which jointly optimizes low rank and TV terms for recovering outlier-free InSAR data stacks. Given the observed InSAR data tensor $ \mathcal{G} $, it can be decomposed into two parts: a low rank tensor $ \mathcal{X} $ and a sparse outlier tensor $ \mathcal{E} $. To maintain smoothness structure of InSAR stacks, the decomposition can be regularized by a TV term. Correspondingly, the proposed total variation regularized robust low rank tensor decomposition method is described by:
\begin{equation}
\begin{aligned}
\{\hat{\mathcal{X}},\hat{\mathcal{E}}\}=&\argmin_{\mathcal{X},\mathcal{E}}\alpha\|\mathcal{X}\|_{3DTV}+\beta\|\mathcal{X}\|_*+\gamma\|\mathcal{E}\|_{1}\\& s.t. \quad \mathcal{G}=\mathcal{X}+\mathcal{E},
\end{aligned}
\label{eq:3DTVLR_orig}
\end{equation}
where $ \|\mathcal{X}\|_{3DTV} $ is the 3D TV term for the three-mode tensor, $ \|\mathcal{X}\|_* $ denotes the tensor nuclear norm, $ \|\mathcal{E}\|_{1} $ is the tensor $ L_1 $ norm of sparse outliers and $ \alpha $, $ \beta $ and $ \gamma $ are the associated parameters for controlling the balance of the three terms. $ \|\mathcal{X}\|_* $ can be calculated by the sum of the $ N $ nuclear norms of the mode-$ n $ unfoldings of $ \mathcal{X} $, i.e. $ \|\mathcal{X}\|_*=\sum_n\|\mathbf{X}_{(n)}\|_* $. The 3D TV term can be defined as:
\begin{equation}
\begin{aligned}
\|\mathcal{X}\|_{3DTV}:=&\sum_{i_1,i_2,i_3}|x_{i_1,i_2,i_3}-x_{i_1,i_2,i_3-1}|+|x_{i_1,i_2,i_3}-x_{i_1,i_2-1,i_3}|+\\&|x_{i_1,i_2,i_3}-x_{i_1-1,i_2,i_3}|.
\end{aligned}
\end{equation}

\subsection{Optimization by Alternating Direction Method of Multipliers (ADMM)}
In order to solve the optimization problem with a TV term, we first introduce auxiliary variables $ \mathcal{Z} $ and $ \mathcal{F} $, and rewrite \eqref{eq:3DTVLR_orig} as:
\begin{equation}
\begin{aligned}
\{\hat{\mathcal{X}},\hat{\mathcal{E}}\}=&\argmin_{\mathcal{X},\mathcal{E}}\alpha\|\mathcal{F}\|_1+\beta\|\mathcal{X}\|_*+\gamma\|\mathcal{E}\|_1\\& s.t. \quad \mathcal{G}=\mathcal{X}+\mathcal{E},\\& \mathcal{X}=\mathcal{Z},~D(\mathcal{Z})=\mathcal{F},
\end{aligned}
\label{eq:3DTVLR}
\end{equation}
where $ D(\cdot)=[D_{i_1}(\cdot);D_{i_2}(\cdot);D_{i_3}(\cdot)]$ is the three-dimensional difference operator and $ D_{i_n}(\cdot) (n=1,2,3) $ is the first-order difference operator with respect to the $ i_n $ dimension of InSAR data stack.

The optimization problem \eqref{eq:3DTVLR} can be solved by the framework of Alternating Direction Method of Multipliers (ADMM) \cite{boyd2011distributed,hong2019cospace,hong2019learnable}. The corresponding constraint optimization problem can be converted into an augmented Lagrangian function, yielding
\begin{equation}
\begin{aligned}
L(&\mathcal{X},\mathcal{E},\mathcal{F},\mathcal{Z},\mathcal{T}_1,\mathcal{T}_2,\mathcal{T}_3)=\alpha\|\mathcal{F}\|_1+\beta\|\mathcal{X}\|_*+\gamma\|\mathcal{E}\|_1+\\&\langle\mathcal{T}_1,\mathcal{G}-\mathcal{X}-\mathcal{E}\rangle+\langle\mathcal{T}_2,\mathcal{X}-\mathcal{Z}\rangle+\langle\mathcal{T}_3,D(\mathcal{Z})-\mathcal{F}\rangle+\\&\frac{\mu}{2}(\|\mathcal{G}-\mathcal{X}-\mathcal{E}\|_F^2+\|\mathcal{X}-\mathcal{Z}\|_F^2+\|D(\mathcal{Z})-\mathcal{F}\|_F^2),
\end{aligned}
\end{equation}
where $ \mathcal{T}_1,\mathcal{T}_2,\mathcal{T}_3 $ are the introduced dual variables and $ \mu $ is the penalty parameter. ADMM takes advantage of splitting one difficult optimization problem into several subproblems, where each of them has a closed-form solution. Accordingly, the minimization of $ L(\mathcal{X},\mathcal{E},\mathcal{F},\mathcal{Z},\mathcal{T}_1,\mathcal{T}_2,\mathcal{T}_3) $ with respect to each variable can be solved by optimizing the following subproblems:
\subsubsection{$ \mathcal{X} $ subproblem}
By fixing the other variables, the subproblem of $ L $ with respect to $ \mathcal{X} $ is:
\begin{equation}
\begin{aligned}
\min_{\mathcal{X}}\beta\|\mathcal{X}\|_*+\frac{\mu}{2}\|\mathcal{X}-\frac{1}{2}(\mathcal{G}-\mathcal{E}+\mathcal{Z}+\frac{\mathcal{T}_1-\mathcal{T}_2}{\mu})\|_F^2.
\end{aligned}
\end{equation}
It can be solved by the Singular Value Thresholding (SVT) operator \cite{cai2010singular,gandy2011tensor} on the mode-$ n (n=1,2,3) $ unfolding of the tensor $ \frac{1}{2}(\mathcal{G}-\mathcal{E}+\mathcal{Z}+\frac{\mathcal{T}_1-\mathcal{T}_2}{\mu}) $, where SVT operator is defined as $ \mathcal{S}_\mu(\mathbf{A}):=\mathbf{U}\mathrm{diag}(\max(\sigma_i-\mu,0))\mathbf{V} $ with $ \mathbf{U} $, $ \mathbf{V} $ and $ \sigma_i $ obtained from Singular Value Decomposition (SVD) of the matrix $ \mathbf{A} $.
\subsubsection{$ \mathcal{Z} $ subproblem}
By fixing the other variables, the subproblem of $ L $ with respect to $ \mathcal{Z} $ has the following form:
\begin{equation}
\begin{aligned}
\min_{\mathcal{Z}}&\langle\mathcal{T}_2,\mathcal{X}-\mathcal{Z}\rangle+\langle\mathcal{T}_3,D(\mathcal{Z})-\mathcal{F}\rangle+\\&\frac{\mu}{2}(\|\mathcal{X}-\mathcal{Z}\|_F^2+\|D(\mathcal{Z})-\mathcal{F}\|_F^2).
\end{aligned}
\end{equation}
Then, by calculating the gradient of $ L $ with respect to $ \mathcal{Z} $ and setting it as zero, we have:
\begin{equation}
	(\mu\mathbf{I}+\mu{D^*}D)\mathcal{Z}=\mathcal{T}_2-D^*(\mathcal{T}_3)+\mu\mathcal{X}+\mu{D^*}(\mathcal{F}),
\end{equation}
where $ D^*(\cdot) $ is the adjoint operator of $ D(\cdot) $. According to the block-circulant structure of the matrix $ {D^*}D $, this inverse problem can be efficiently solved by exploiting 3D Fast Fourier Transform (FFT) and its inverse transform \cite{yao2017tvlowrank,ji2016tensor}. 

\subsubsection{$ \mathcal{F} $ subproblem}
By fixing the other variables, the subproblem of $ L $ with respect to $ \mathcal{F} $ can be written as:
\begin{equation}
\begin{aligned}
\min_{\mathcal{F}}\alpha\|\mathcal{F}\|_1+\frac{\mu}{2}\|\mathcal{F}-D(\mathcal{Z})-\frac{\mathcal{T}_3}{\mu}\|_F^2.
\end{aligned}
\end{equation}
This $ L_1 $-norm-induced subproblem can be efficiently solved by applying the soft-thresholding operator defined as $ \mathcal{R}_\gamma(\mathcal{A}):=\mathrm{sign}(\mathcal{A})\odot\max(|\mathcal{A}|-\gamma,0) $, where $ \odot $ denotes the element-wise product (Hadamard
product) of two tensors, and $ |\mathcal{A}|=\mathrm{sign}(\mathcal{A})\odot\mathcal{A}$.

\subsubsection{$ \mathcal{E} $ subproblem}
By fixing the other variables, the subproblem of $ L $ with respect to $ \mathcal{E} $ is:
\begin{equation}
\begin{aligned}
\min_{\mathcal{E}}\gamma\|\mathcal{E}\|_1+\frac{\mu}{2}\|\mathcal{E}-\mathcal{G}+\mathcal{X}-\frac{\mathcal{T}_1}{\mu}\|_F^2.
\end{aligned}
\end{equation}
Likewise, this subproblem can also be solved by soft-thresholding operator.

\subsubsection{Multiplier Updating}
All the dual variables can be updated by:
\begin{equation}
\begin{aligned}
&\mathcal{T}_1=\mathcal{T}_1+\mu(\mathcal{G}-\mathcal{X}-\mathcal{E}),\\&
\mathcal{T}_2=\mathcal{T}_2+\mu(\mathcal{X}-\mathcal{Z}),\\&
\mathcal{T}_3=\mathcal{T}_3+\mu(D(\mathcal{Z})-\mathcal{F}).
\end{aligned}
\end{equation}

The detailed ADMM pseudocode for solving \eqref{eq:3DTVLR} is summarized in Algorithm \ref{ag:ADMM_solver}.

Using a predefined convergence condition, the solution ($ \hat{\mathcal{X}} $ and $ \hat{\mathcal{E}} $) can be obtained, i.e. the outlier-free InSAR data tensor and the sparse outlier tensor, respectively. To this end, by applying conventional multipass InSAR techniques, e.g. PSI \cite{ferretti2001permanent}, on $ \hat{\mathcal{X}} $, we can robustly retrieve the geophysical parameters.

\section{Simulations}\label{sc:simulation}

\subsection{Simulation Results}
\begin{figure*}
	\centering
	\includegraphics[width=\textwidth]{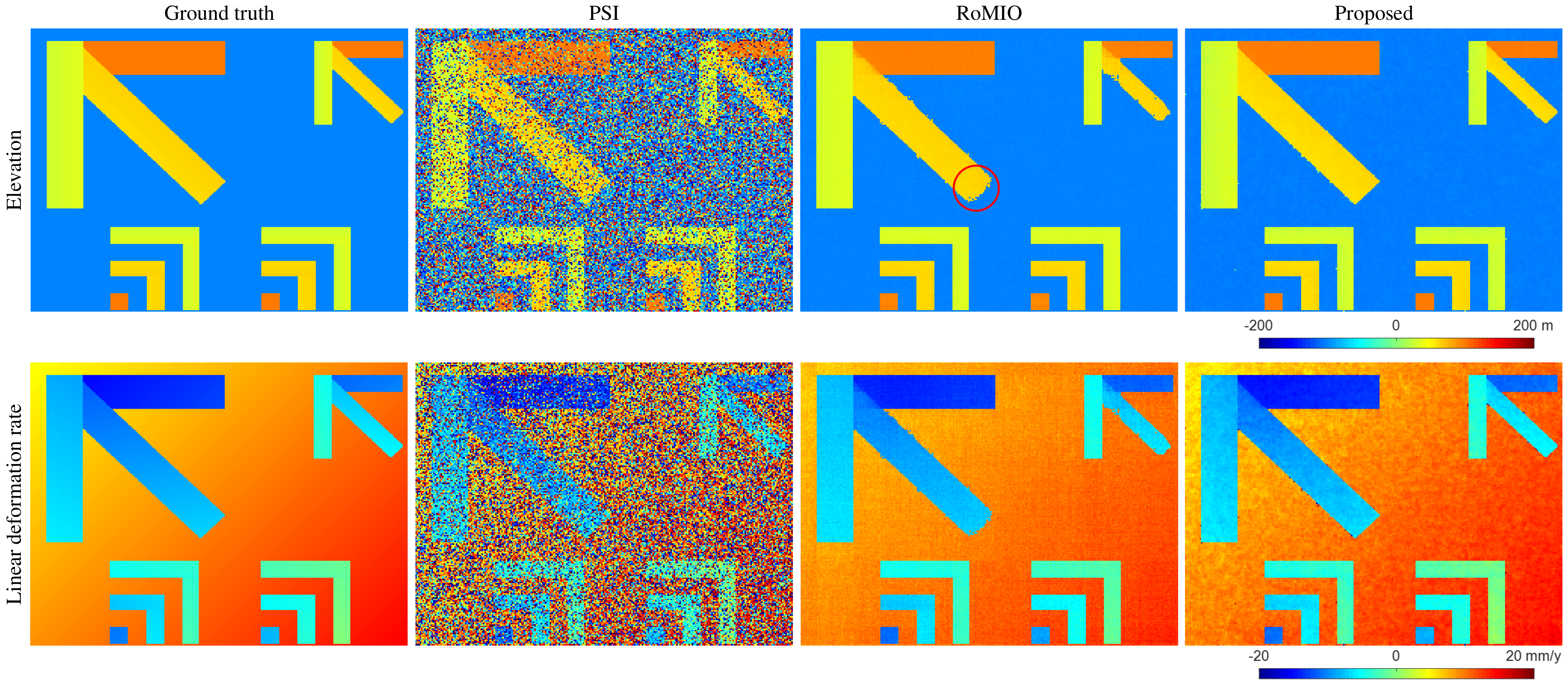}
	\caption{The simulated ground truth maps of linear deformation rate and elevation, along with the estimated results by PSI, RoMIO \cite{kang2018romio} and the proposed method. Uncorrelated complex circular Gaussian noise was added to the simulated InSAR stack with an SNR of $ 0 $dB, i.e. according to PS model. To simulate sparse outliers in the stacks, $ 20\% $ of pixels randomly selected from the stack were replaced with uniformly distributed phases. It can be seen that most points cannot be correctly estimated by PSI. Especially for the estimates of the ground deformation, the increasing trend from top left to the bottom right corner is not clearly visible in the PSI result. As a comparison, both the patterns of elevation and deformation maps from RoMIO and the proposed method are more clearly displayed than PSI. However, without TV regularization, the reconstruction of some "building blocks" is more blurred in RoMIO than the proposed method, e.g. the area indicated by the red circle.}
	\label{fg:simu_reso_results}
\end{figure*}

We simulated a multipass InSAR data stack of $ 200\times250 $ pixels by $ 29 $ images with the true elevation and linear deformation rate shown in \Fig \ref{fg:simu_reso_results}. The simulation is comparable to real scenario of urban areas. The flat background of the elevation map and different blocks on it represent the ground and buildings with different heights, respectively. Also, as shown by the simulated deformation map, gradually increasing displacement is often observed in real data. The linear deformation rates range from $ -15 \mathrm{mm/year} $ to $ 15 \mathrm{mm/year} $ and elevations are from $ -100 \mathrm{m} $ to $ 100 \mathrm{m} $. The spatial baseline and the temporal baseline were chosen to be comparable to those of TerraSAR-X. Uncorrelated complex circular Gaussian noise was added to the simulated stack with an signal-to-noise ratio (SNR) of $ 0 $dB i.e. following the PS model. To simulate sparse outliers in the stacks, $ 20\% $ of pixels randomly selected from the stack were replaced with uniformly distributed phases. 

\begin{figure}
	\centering
	\includegraphics[width=0.24\textwidth]{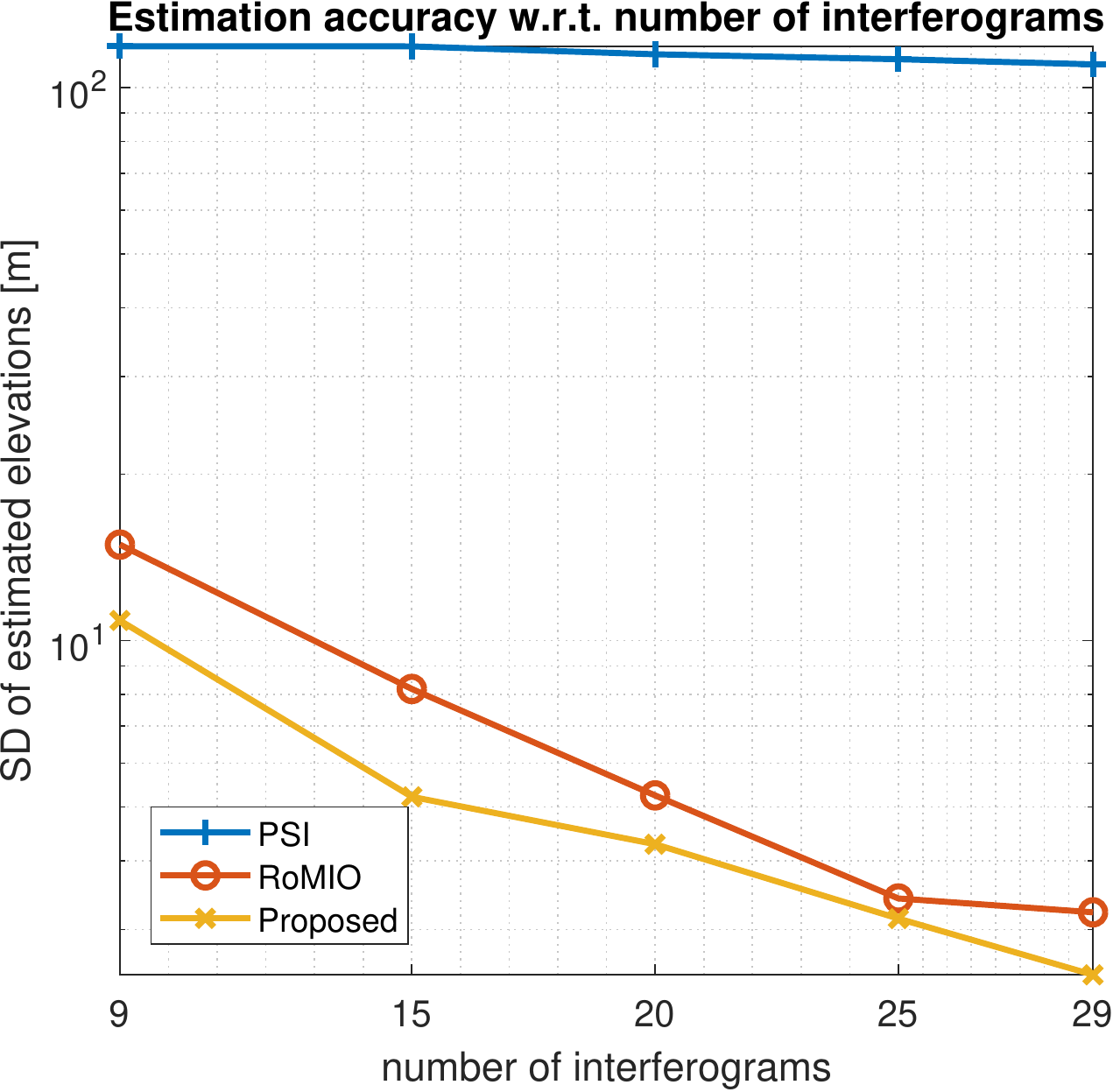}~
	\includegraphics[width=0.24\textwidth]{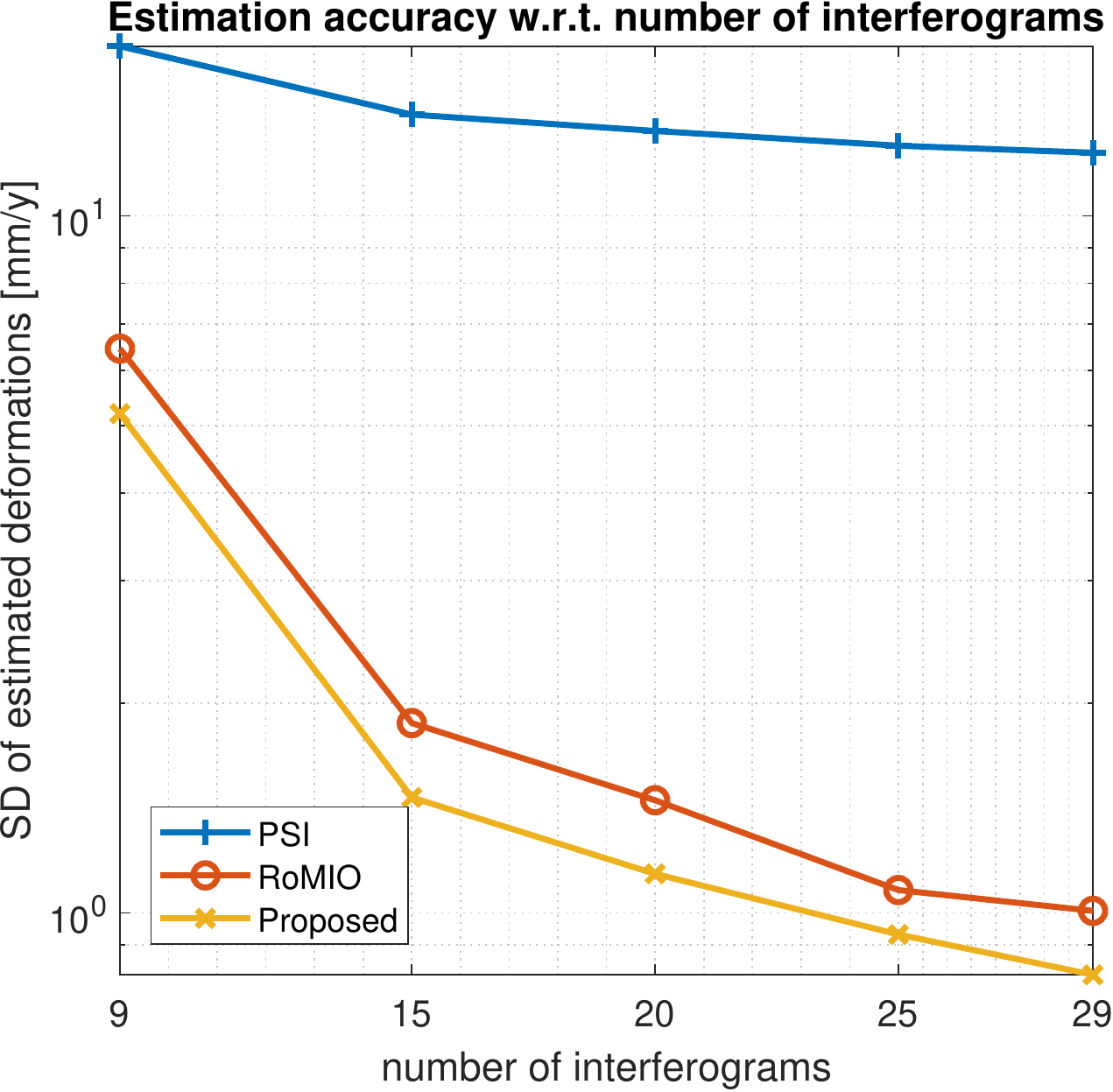}
	\caption{Plot of the estimation accuracy with respect to different numbers of interferograms. As the number of interferograms utilized for the reconstruction decreases, the performances of all the methods decline, but our method can still maintain the best enhancement of the estimation accuracy.}
	\label{fg:SD_wrt_IMG_NUM}
\end{figure}

\begin{figure}
	\centering
	\includegraphics[width=0.24\textwidth]{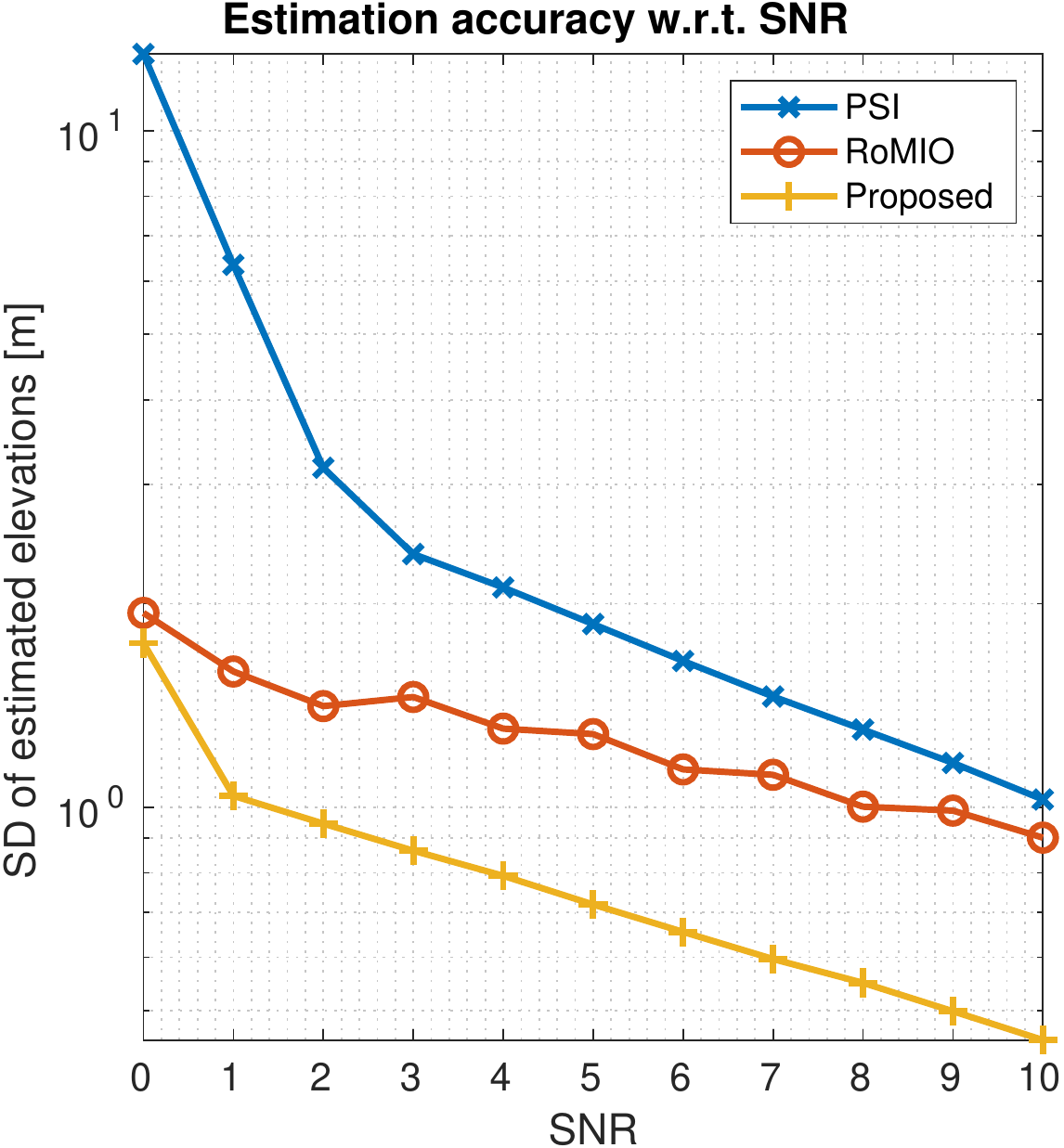}~
	\includegraphics[width=0.24\textwidth]{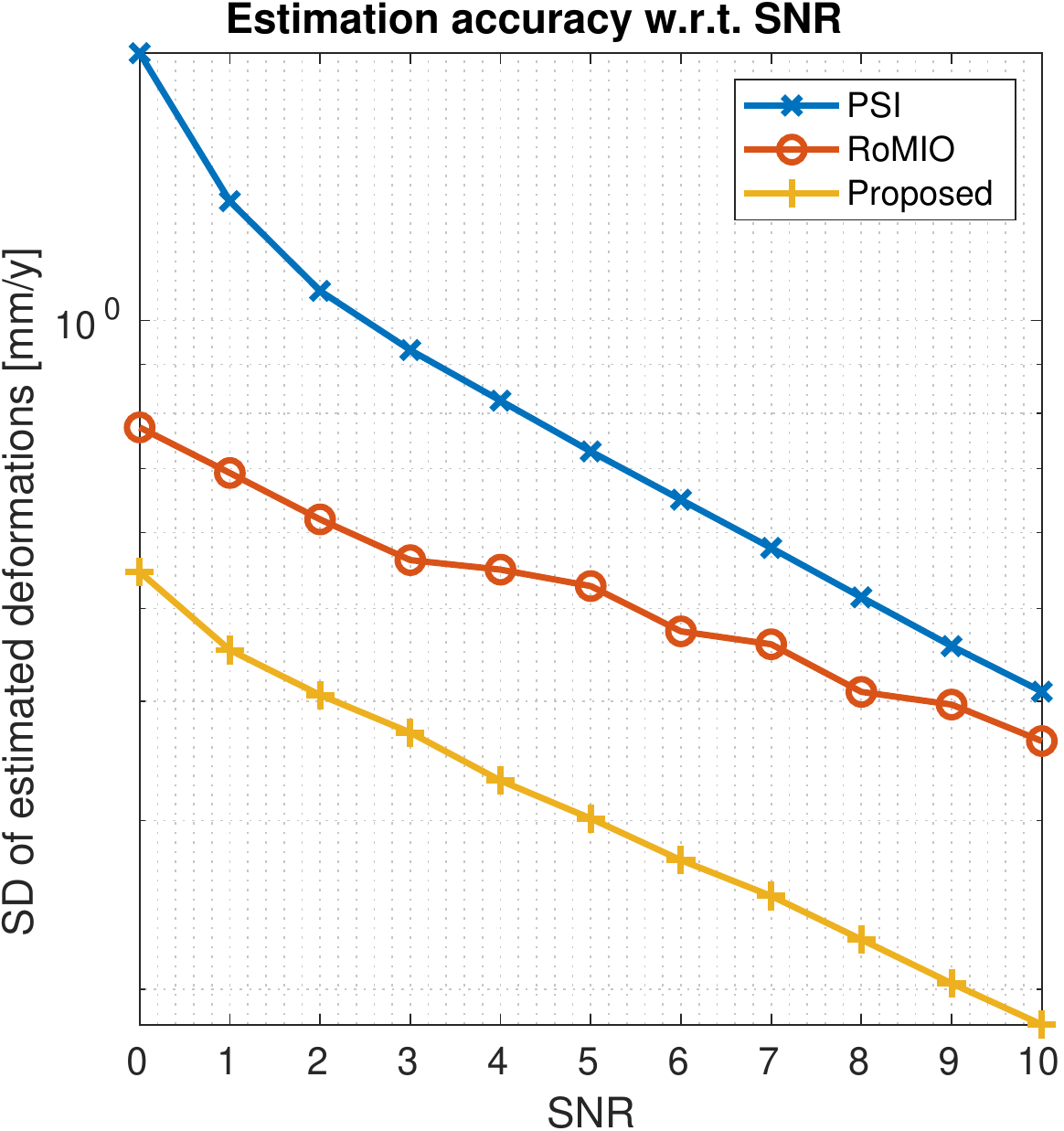}~
	\caption{Plot of the estimation accuracy with respect to different values of SNR. As SNR grows, the efficiency improvement of RoMIO is less prominent than the proposed method. One plausible reason may be owing to the mitigation effect of Gaussian noise by TV regularization in the proposed method.}
	\label{fg:SD_wrt_SNR}
\end{figure}

\begin{figure}
	\centering
	\includegraphics[width=0.24\textwidth]{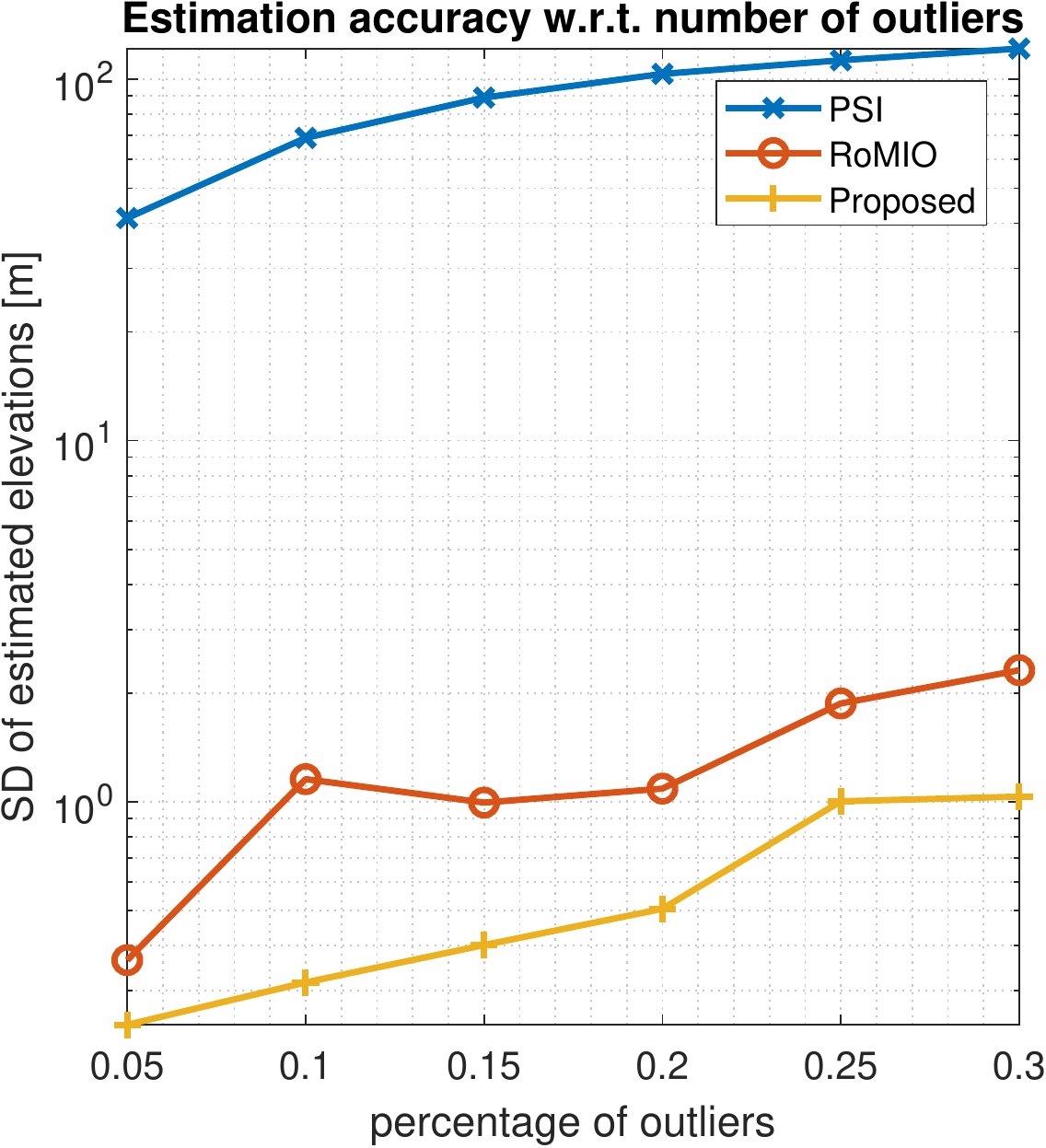}~
	\includegraphics[width=0.24\textwidth]{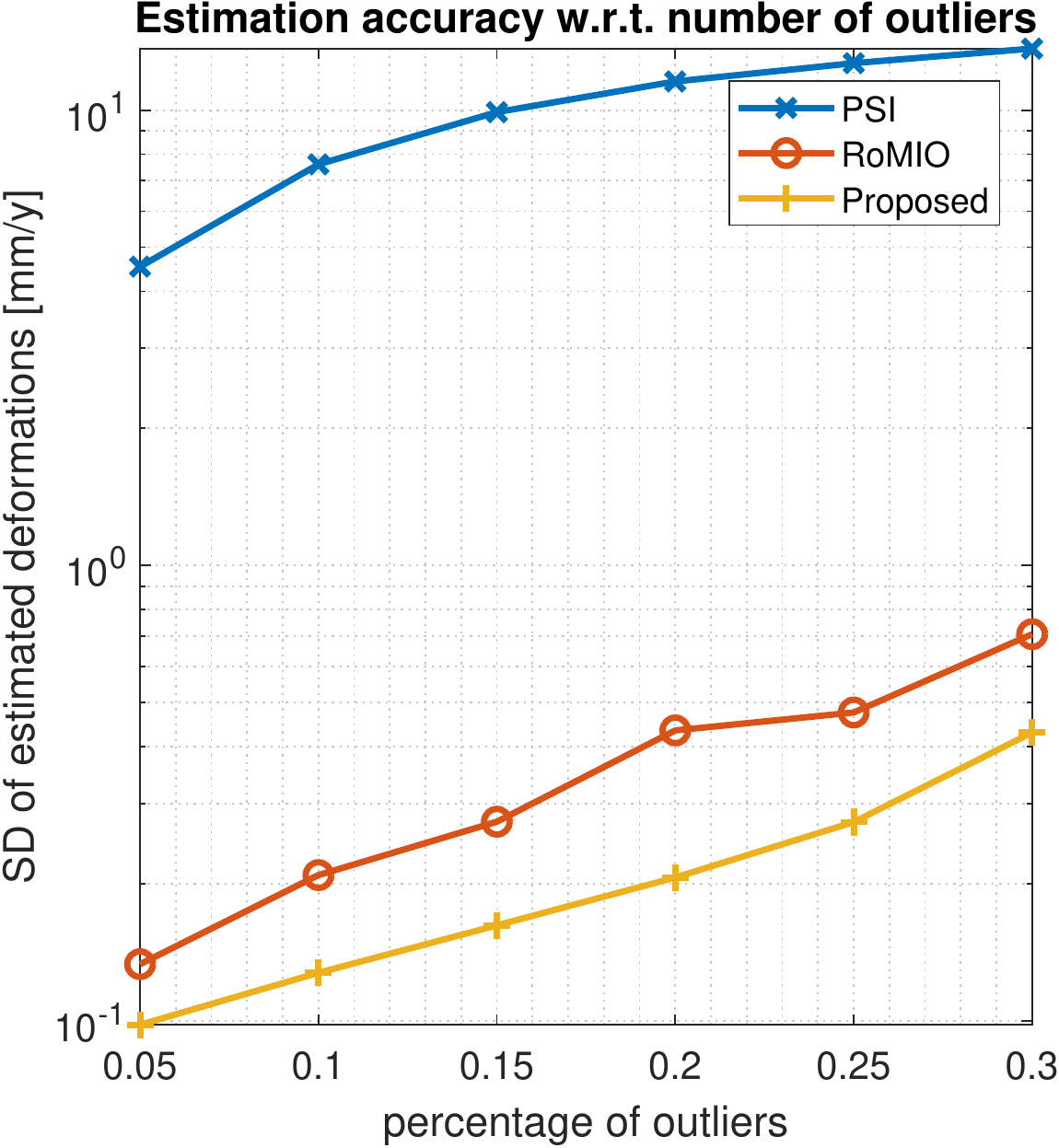}
	\caption{Plot of the estimation accuracy with respect to different percentages of outliers. It can be seen that both RoMIO and the proposed method can robustly estimate geophysical parameters.}
	\label{fg:SD_wrt_out}
\end{figure}

As illustrated in \Fig \ref{fg:simu_reso_results}, we compared the geophysical parameters estimated by PSI, Robust Multipass InSAR technique via Object-based low rank tensor decomposition (RoMIO) \cite{kang2018romio} and the proposed method. The parameters of the proposed method are set as $ \alpha=0.1 $, $ \beta=2 $ and $ \gamma=0.48 $, respectively. The parameter selection is discussed in the following subsection. Furthermore, as shown in \Fig \ref{fg:SD_wrt_IMG_NUM}. in order to test the capability of the proposed method for handling small stacks, we calculated standard deviation (SD) of the residuals between the estimated parameters and the ground truth with respect to decreasing number of interferograms down to $ 9 $. Besides, the performance of the proposed method against different values of SNR and percentages of outliers were tested and plotted in \Fig \ref{fg:SD_wrt_SNR} and \ref{fg:SD_wrt_out}.  

\subsection{Parameter Selection}
\begin{figure}
	\centering
	\includegraphics[width=0.25\textwidth]{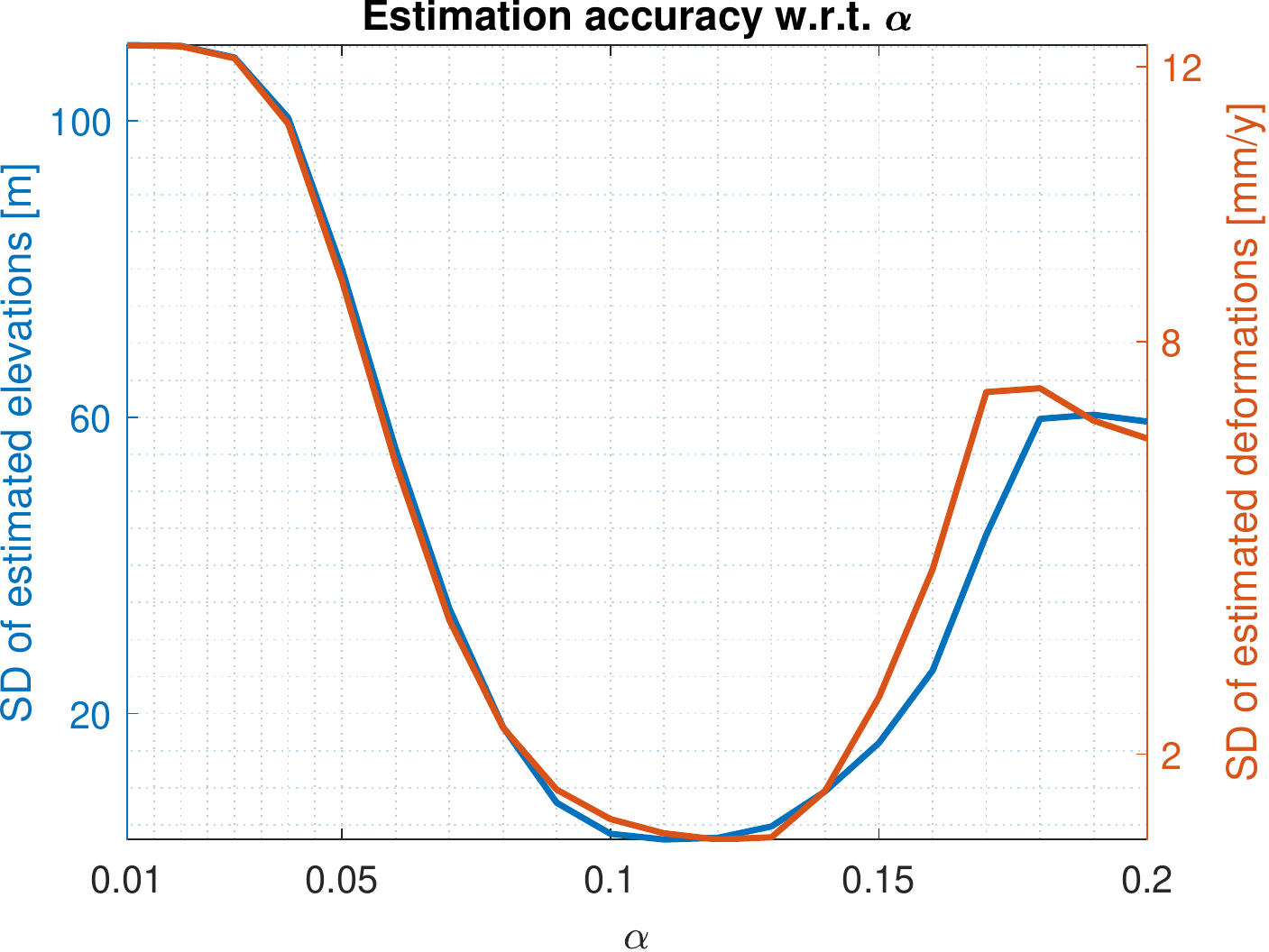}~
	\includegraphics[width=0.24\textwidth]{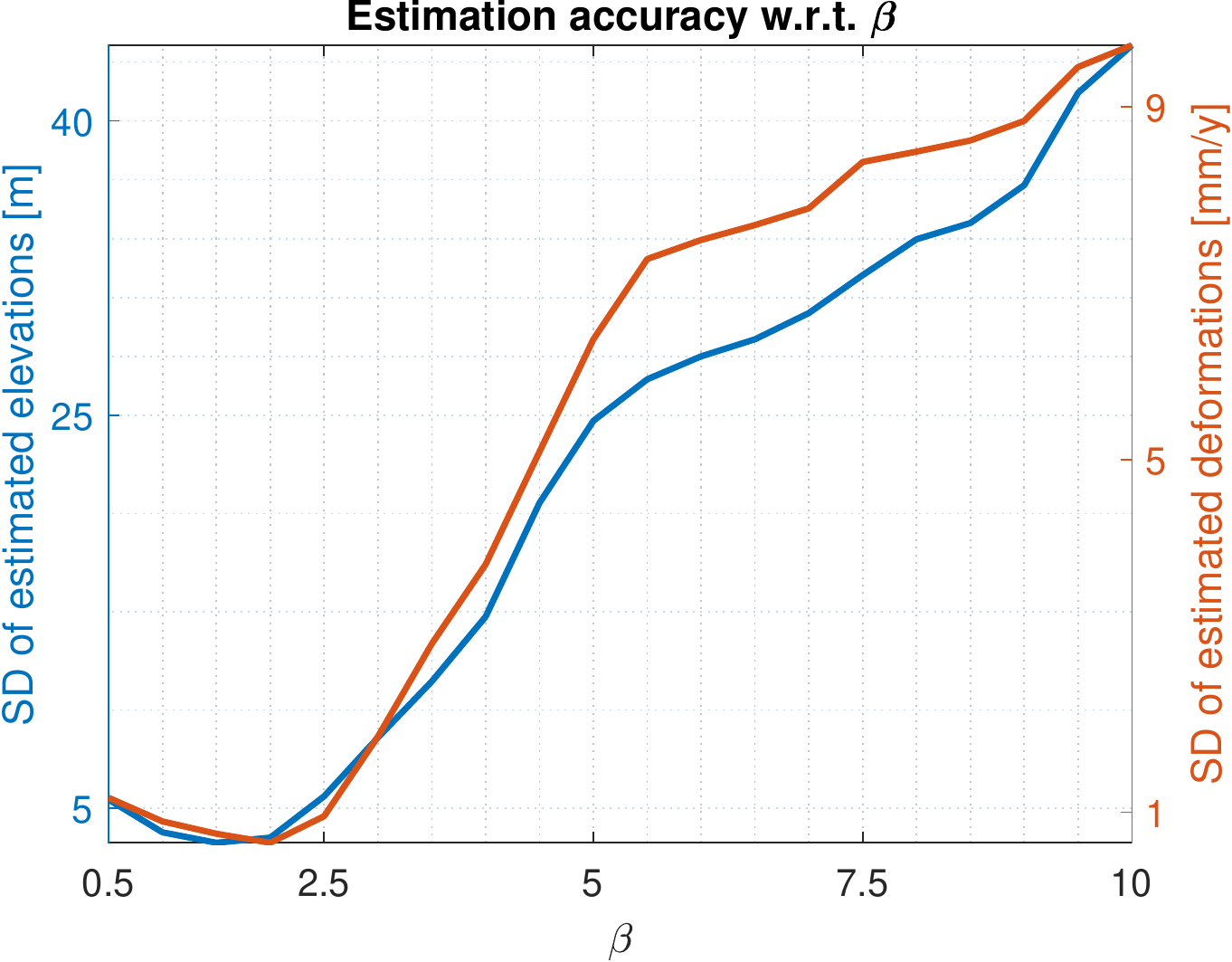}
	\caption{Plot of the estimation accuracy with respect to different parameter settings of $ \alpha $ and $ \beta $. The optimal $ \alpha $ and $ \beta $ for this simulation are around $ 0.11 $ and $ 1 $, respectively. }
	\label{fg:SD_wrt_alpha_beta}
\end{figure}

There are totally four parameters introduced in the proposed method, i.e. $ \alpha,\beta,\gamma,\mu $, where $ \alpha,\beta,\gamma $ control the balances of the three optimization terms and $ \mu $ comes with the Lagrange multiplier terms. $ \mu $ can be initially set as $ 10^{-2} $ and updated in each iteration by $ \mu:=\min(\eta\mu,\mu_{\max}) $, where $ \eta=1.1 $. As introduced in \cite{yao2017tvlowrank,xu2018joint,xu2019nonlocal}, $ \gamma $ can be set as $ 100/\sqrt{I_1I_2} $. In our experiences, $ \alpha $ is selected in a range from $ 0 $ to $ 0.2 $ and $ \beta $ can be chosen between $ 0 $ to $ 10 $. As shown in \Fig \ref{fg:SD_wrt_alpha_beta}, based on the simulation of \Fig \ref{fg:SD_wrt_out}, we performed the estimation accuracy of the parameters with respect to different values of $ \alpha $ and $ \beta $. It can be seen that optimal $ \alpha $ and $ \beta $ for this simulation are around $ 0.11 $ and $ 1 $, respectively.

\subsection{Performance Analysis}
According to the visualization results shown in \Fig \ref{fg:simu_reso_results}, under SNR=$ 0 $dB and $ 20\% $ outliers, most points cannot be correctly estimated by PSI. Especially for the background of deformation map, the increasing trend from top left to the bottom right corner is not clearly visible in the PSI result. As a comparison, both the patterns of elevation and deformation maps from RoMIO and the proposed method are more clearly displayed than PSI. However, without TV regularization, the reconstruction of some "building blocks" is more blurred in RoMIO than the proposed method, e.g. the area indicated by the red circle in \Fig \ref{fg:simu_reso_results}, since piece-wise smoothness cannot be maintained by RoMIO. Besides, as displayed by the deformation results, non-piece-wise smoothness information can also be preserved in the proposed method. As shown in \Fig \ref{fg:SD_wrt_IMG_NUM}, under this simulation, the improvement of the estimation accuracy by both RoMIO and the proposed method can achieve ten times better than PSI. Besides, as the number of interferograms utilized for the reconstruction decreases, the performances of all the methods decline, but our method can still maintain the best estimation accuracy. Based on \Fig \ref{fg:SD_wrt_SNR} and \ref{fg:SD_wrt_out}, we can see that the proposed method can mitigate the influences from both complex Gaussian noises and outliers in the InSAR stack and accomplishes more accurate reconstruction than the other two methods. As SNR grows, the efficiency improvement of RoMIO is less prominent than the proposed method. One plausible reason may be owing to the mitigation effect of Gaussian noise by TV regularization in the proposed method. It can be also observed that the performance of PSI is more severely impacted by outliers than complex Gaussian noise. The reason lies in the fact that Periodogram exploited in PSI are only the Maximum Likelihood (ML) estimator under complex Gaussian noise. It is not robust to outliers.     
\section{Case Study Using Real Data}\label{sc:real_data}
\subsection{Real Data Results}

\begin{figure}
	\centering
	\includegraphics[width=0.5\textwidth]{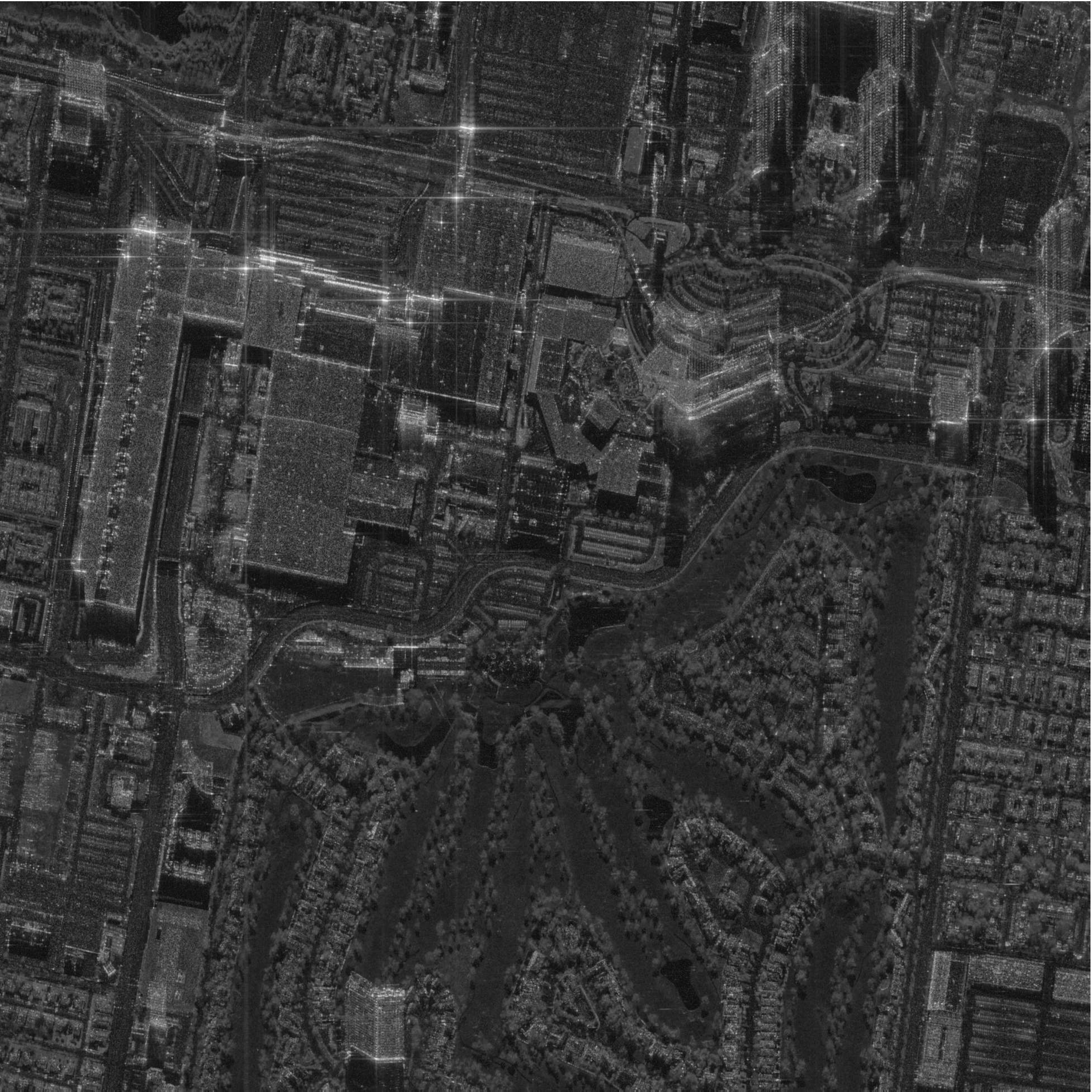}
	\caption{The study area of Las Vegas shown by the mean amplitude (log scale) of a TerraSAR-X InSAR stack.}
	\label{fg:lasvegas_SAR_img}
\end{figure}

\begin{figure}
	\centering
	\includegraphics[width=0.48\textwidth]{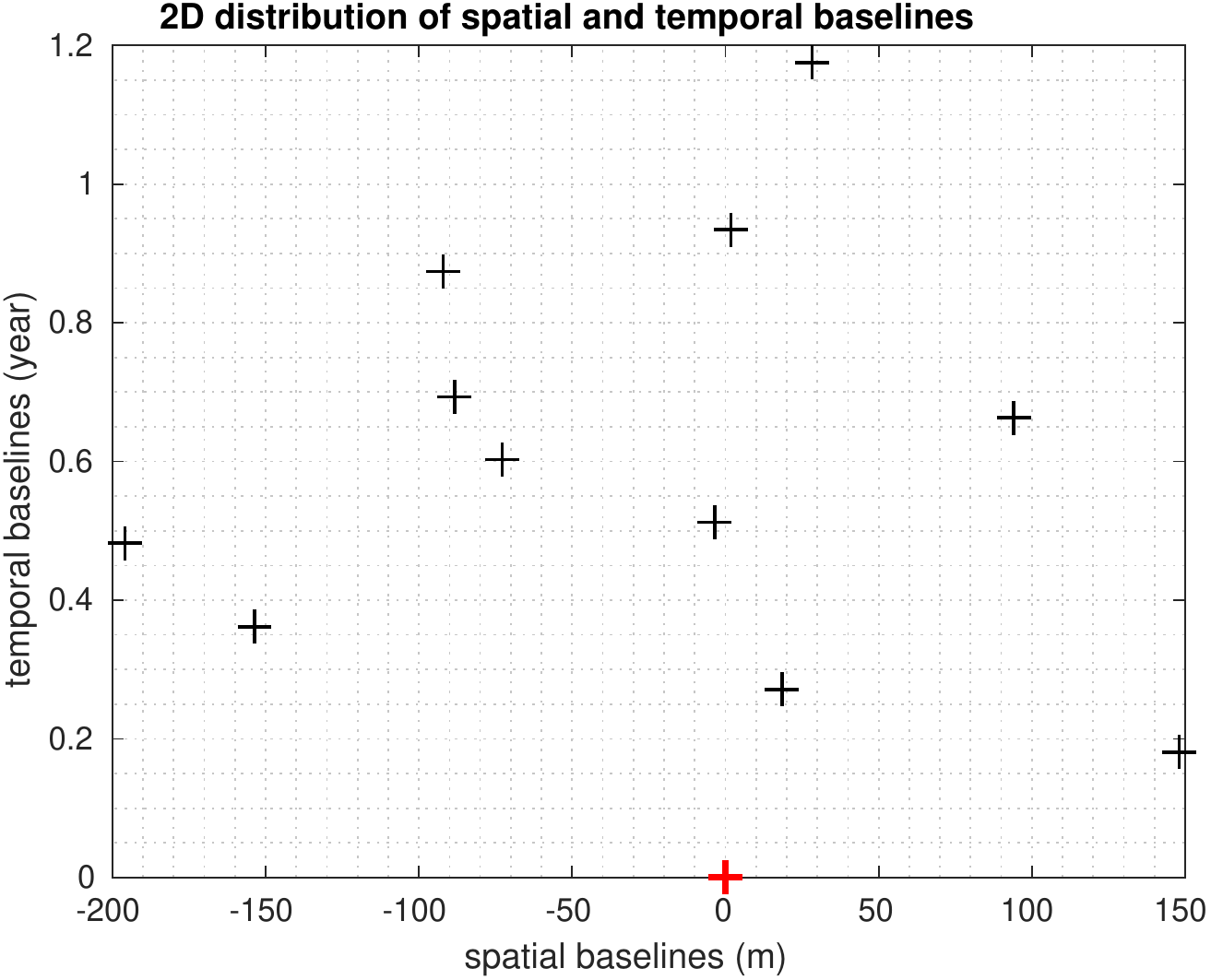}
	\caption{The 2D distribution of spatial and temporal baselines of the selected $ 11 $ interferograms for reconstruction. The master baseline is shown in red.}
	\label{fg:lasvegas_spatial_temp_baseline}
\end{figure}

\begin{figure*}
	\centering
	\includegraphics[width=\textwidth]{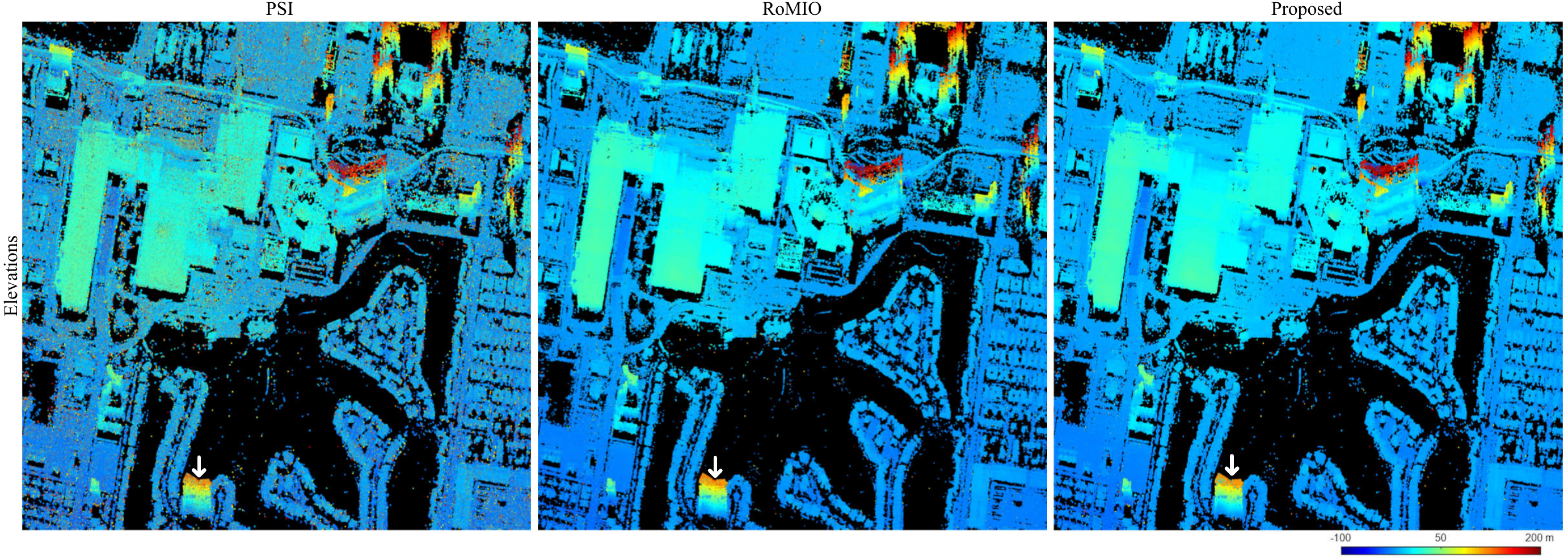}
	\caption{Estimated elevation maps by PSI, RoMIO and the proposed method with $ 11 $ interferograms of one area in Las Vegas. Consistent with the simulations, the tensor-decomposition-based methods, i.e. RoMIO and the proposed method, can achieve more robust performances than PSI, since many noisy points are observed in the result of PSI. For a detailed comparison, profiles of building fa\c{c}ade (indicated by the white arrows) are plotted in \Fig \ref{fg:lasvegas_ele_profile}.}
	\label{fg:lasvegas_ele_map}
\end{figure*}

\begin{figure*}
	\centering
	\includegraphics[width=\textwidth]{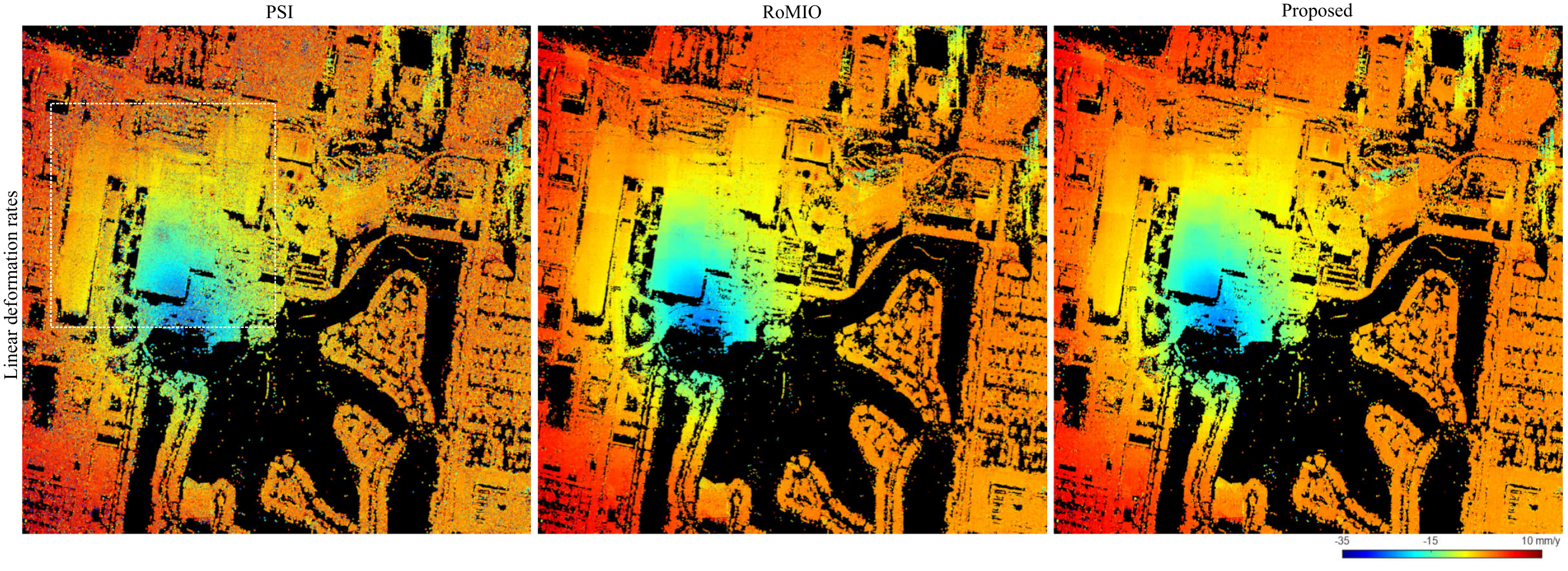}
	\caption{Estimated linear deformation rates by PSI and the proposed method with $ 11 $ interferograms of one area in Las Vegas. Obviously, tensor-decomposition-based methods, RoMIO and the proposed one, can better maintain the smoothness of the reconstructed deformation maps. The reconstruction results of the convention center (white rectangular) are displayed in \Fig \ref{fg:lasvegas_defor_map_zoomin}.}
	\label{fg:lasvegas_defor_map}
\end{figure*}

\begin{figure*}
	\centering
	\includegraphics[width=\textwidth]{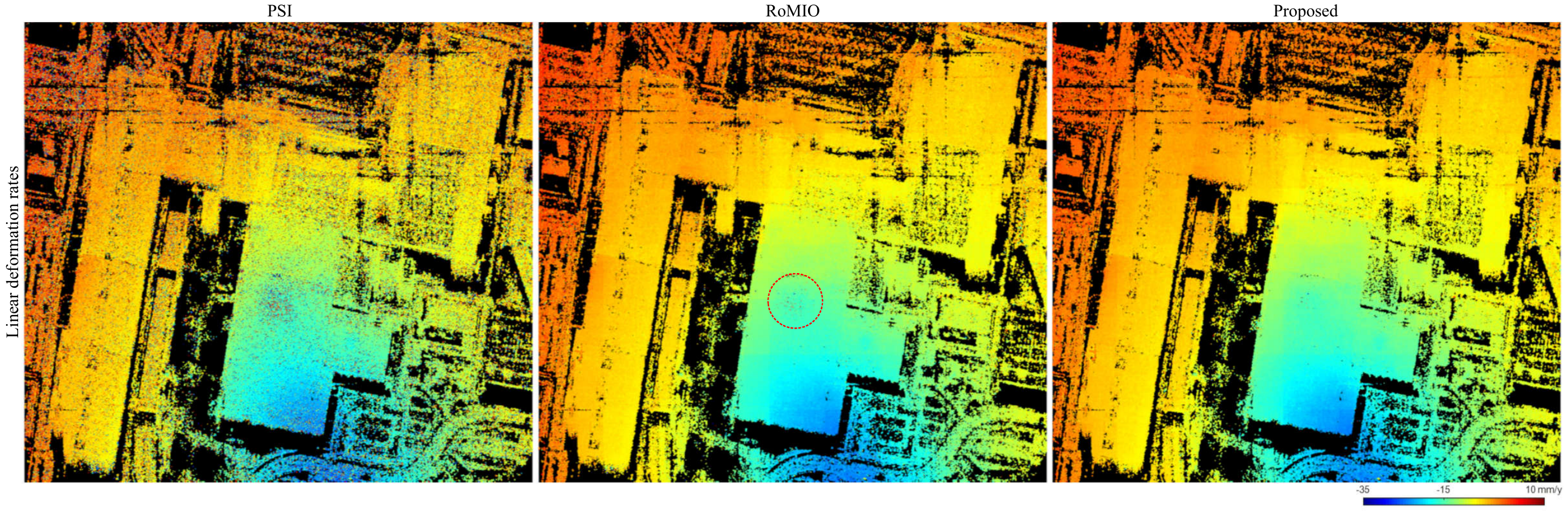}
	\caption{The cropped zoom-in areas of the results in \Fig \ref{fg:lasvegas_defor_map} by the dashed white rectangular. Compared to RoMIO, the proposed method can better estimate the flat roof areas, since the group of noisy points (indicated by red dashed circle) is eliminated in the result of the proposed method. }
	\label{fg:lasvegas_defor_map_zoomin}
\end{figure*}

\subsubsection{Las Vegas}
The first study area is in Las Vegas, as demonstrated in \Fig \ref{fg:lasvegas_SAR_img}. The InSAR stack contains $ 29 $ TerraSAR-X interferograms in total, with the spatial dimension of $ 1950\times1950 $ pixels. In order to test the performance of the proposed method under a low number of interferograms, a substack with $ 11 $ interferograms were selected from the full stack. The interferograms were selected so that their spatial and temporal baselines are close to uniform distribution, which is illustrated in \Fig \ref{fg:lasvegas_spatial_temp_baseline}. Since this spatial area is relatively large, RoMIO and the proposed method were conducted in a sliding-window manner, with a patch size of $ 100\times100 $ pixels. The parameters of our method were set as $ \alpha=0.12 $, $ \beta=5 $ and $ \gamma=1 $. The estimated elevations and linear deformation rates by PSI, RoMIO and the proposed method are displayed in \Fig \ref{fg:lasvegas_ele_map} and \ref{fg:lasvegas_defor_map}, respectively.

\subsubsection{Berlin}
\begin{figure}
	\centering
	\includegraphics[width=0.5\textwidth]{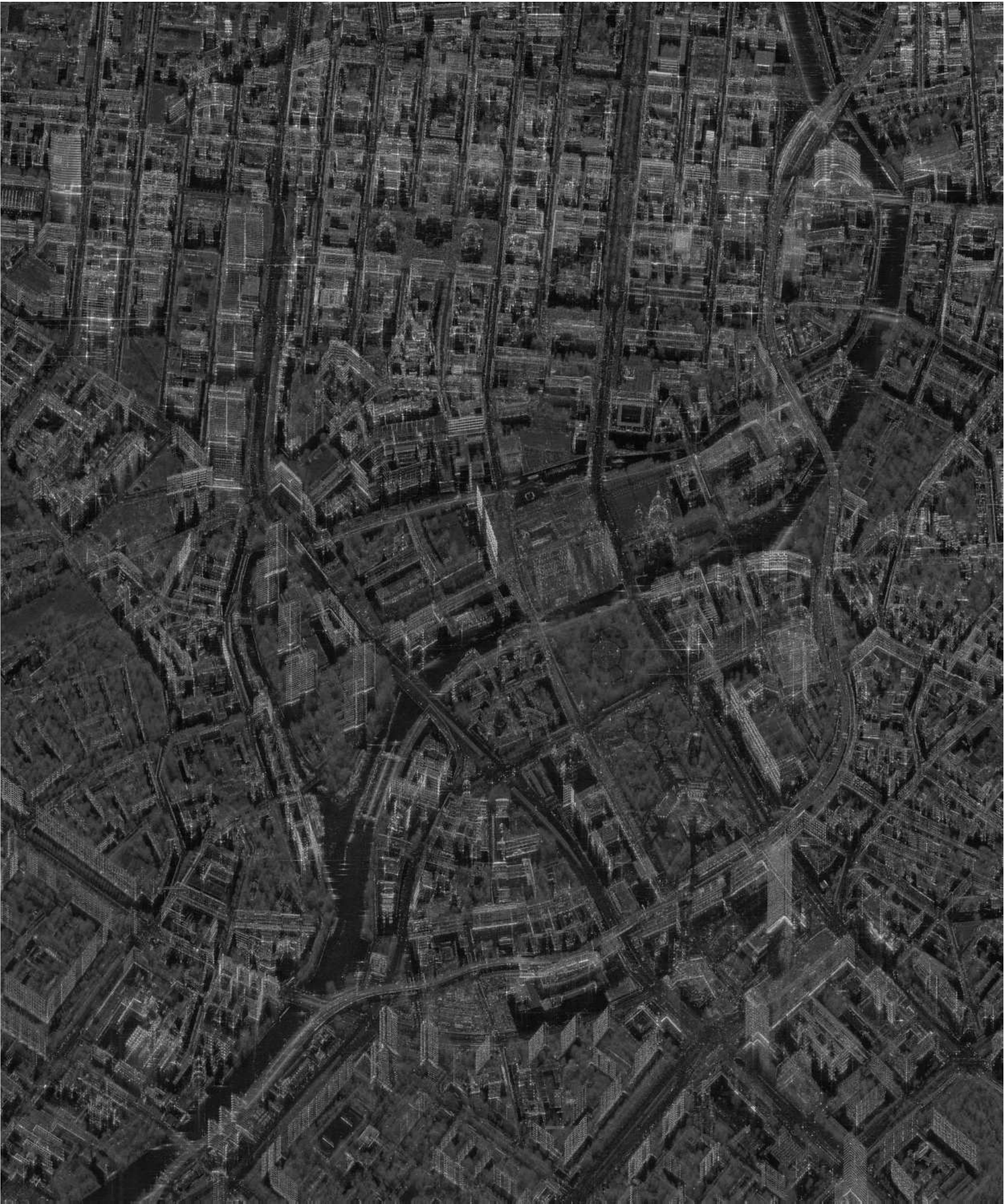}
	\caption{The study area of Berlin shown by the mean amplitude (log scale) of a TerraSAR-X InSAR stack.}
	\label{fg:berlin_SAR_img}
\end{figure}

\begin{figure}
	\centering
	\includegraphics[width=0.5\textwidth]{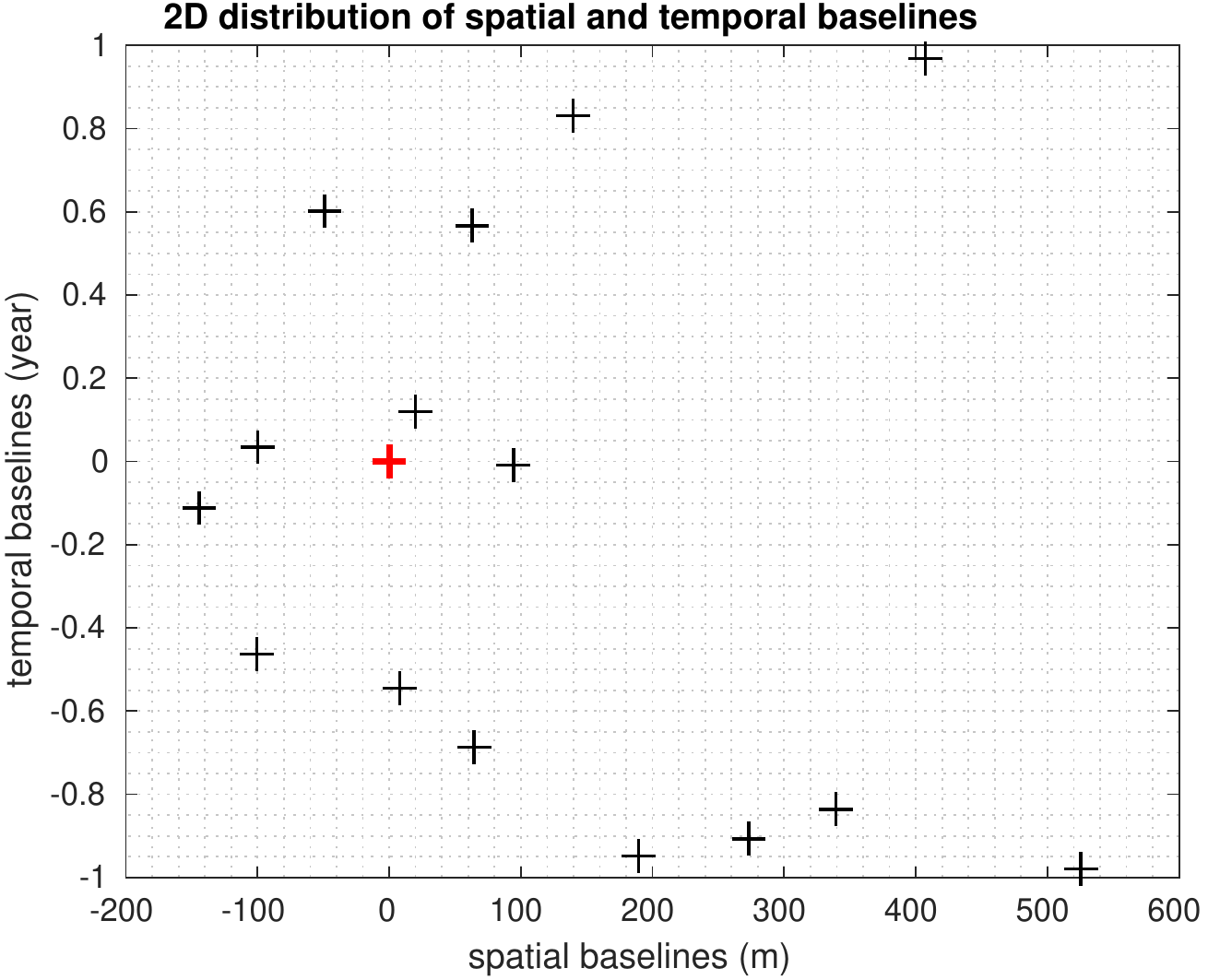}
	\caption{The 2D distribution of spatial and temporal baselines of the selected $ 15 $ interferograms for reconstruction. The master baseline is shown in red.}
	\label{fg:berlin_spatial_temp_baseline}
\end{figure}

\begin{figure*}
	\centering
	\includegraphics[width=\textwidth]{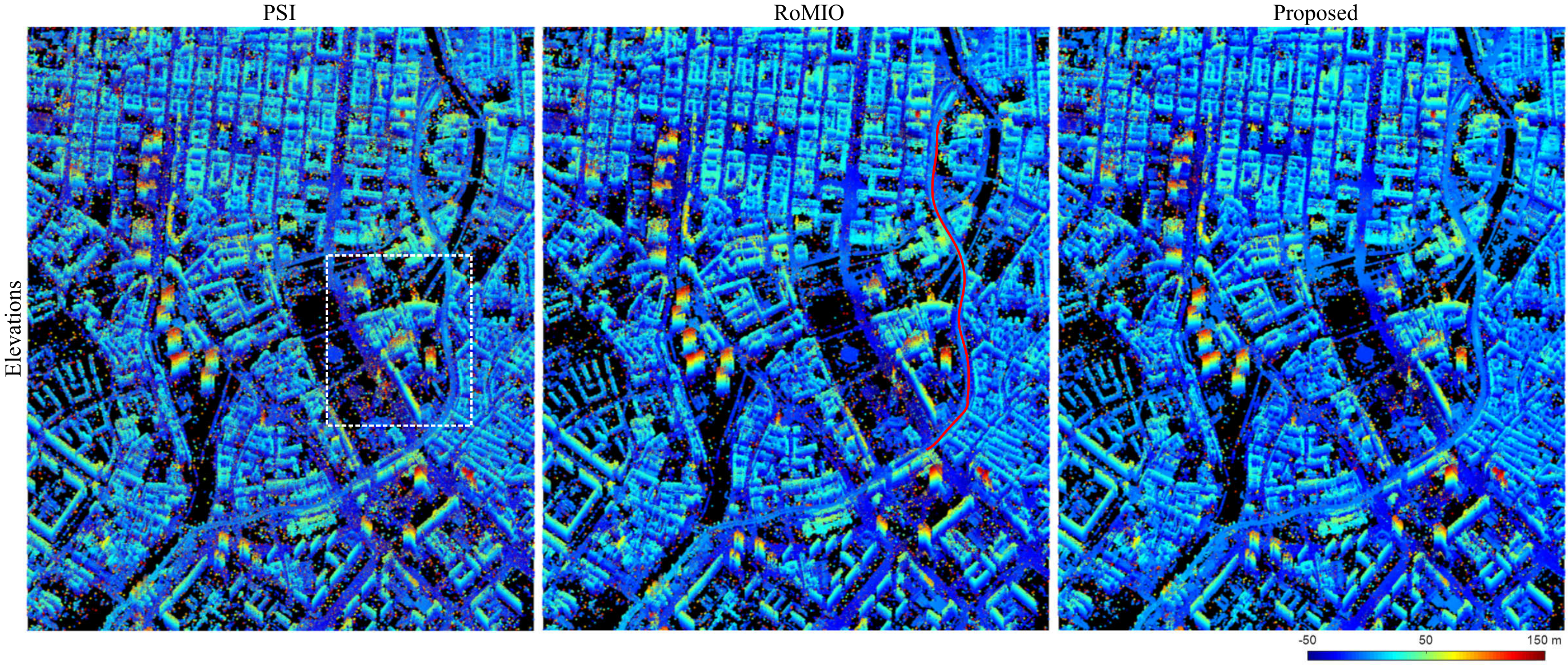}
	\caption{Estimated elevation maps by PSI, RoMIO and the proposed method with $ 15 $ interferograms of one area in Berlin. Besides the reconstruction of flat areas as Las Vegas, the proposed method can also achieve the robust retrieval of this complex area composed by building blocks and high-rise buildings. For a better comparison of the three methods, one zoom-in area and one road profile are displayed in \Fig \ref{fg:berlin_ele_map_zoomin} and \ref{fg:berlin_road_profile}, respectively.}
	\label{fg:berlin_ele_map}
\end{figure*}

\begin{figure*}
	\centering
	\includegraphics[width=\textwidth]{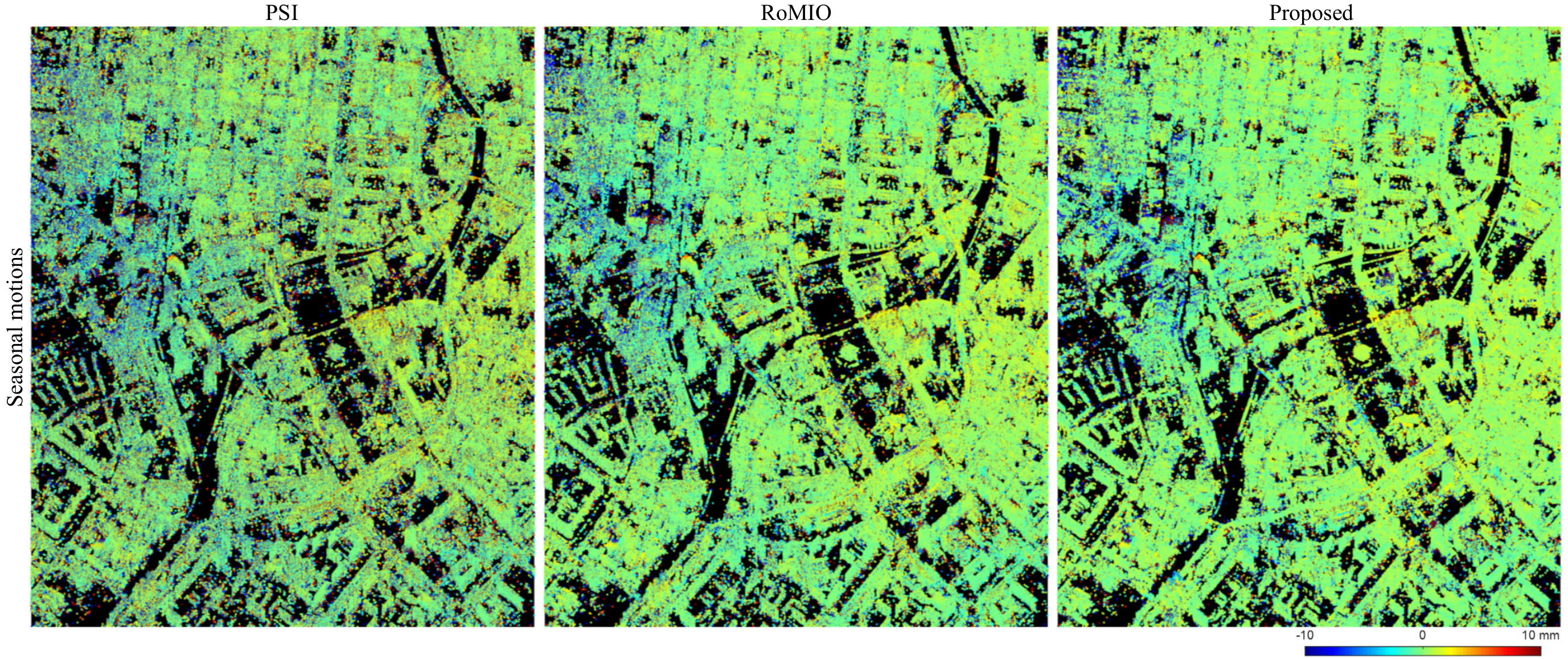}
	\caption{Estimated amplitudes of seasonal motions by PSI, RoMIO and the proposed method with $ 15 $ interferograms of one area in Berlin. Smoothness structure can be well maintained in the reconstructed deformation map by the proposed method.}
	\label{fg:berlin_defor_map}
\end{figure*}

Another study area is in Berlin, as shown in \Fig \ref{fg:berlin_SAR_img}. The InSAR stack totally contains in total $ 41 $ TerraSAR-X interferograms, with the spatial dimension of $ 3000\times2500 $ pixels. A substack with $ 15 $ interferograms were selected from the full stack and the associated baselines were plotted in \Fig \ref{fg:berlin_spatial_temp_baseline}. Likewise, the patch size used in the sliding-window processing is chosen as $ 200\times200 $ pixels. For this area, the parameters of our method were set as $ \alpha=0.12 $, $ \beta=3 $ and $ \gamma=0.5 $. The estimated elevations and amplitudes of seasonal motions by PSI, RoMIO and the proposed method are displayed in \Fig \ref{fg:berlin_ele_map} and \ref{fg:berlin_defor_map}, respectively.

\begin{figure*}
	\centering
	\includegraphics[width=\textwidth]{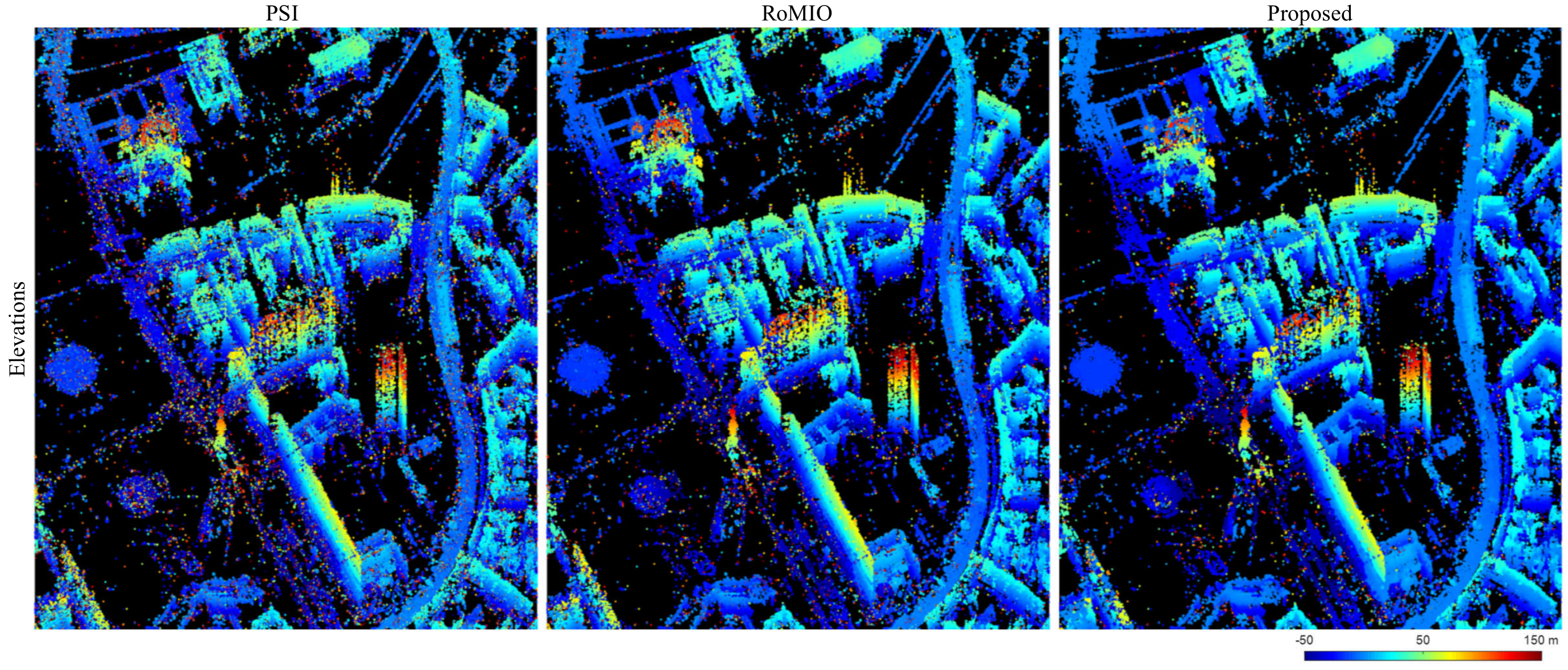}
	\caption{The cropped zoom-in areas of the results in \Fig \ref{fg:berlin_ele_map} by the dashed white rectangular. Compared to PSI, most outliers can be mitigated by the tensor-decomposition-based methods.}
	\label{fg:berlin_ele_map_zoomin}
\end{figure*}

\subsection{Performance Analysis}

\subsubsection{Las Vegas}
\begin{figure}
	\centering
	\includegraphics[width=0.5\textwidth]{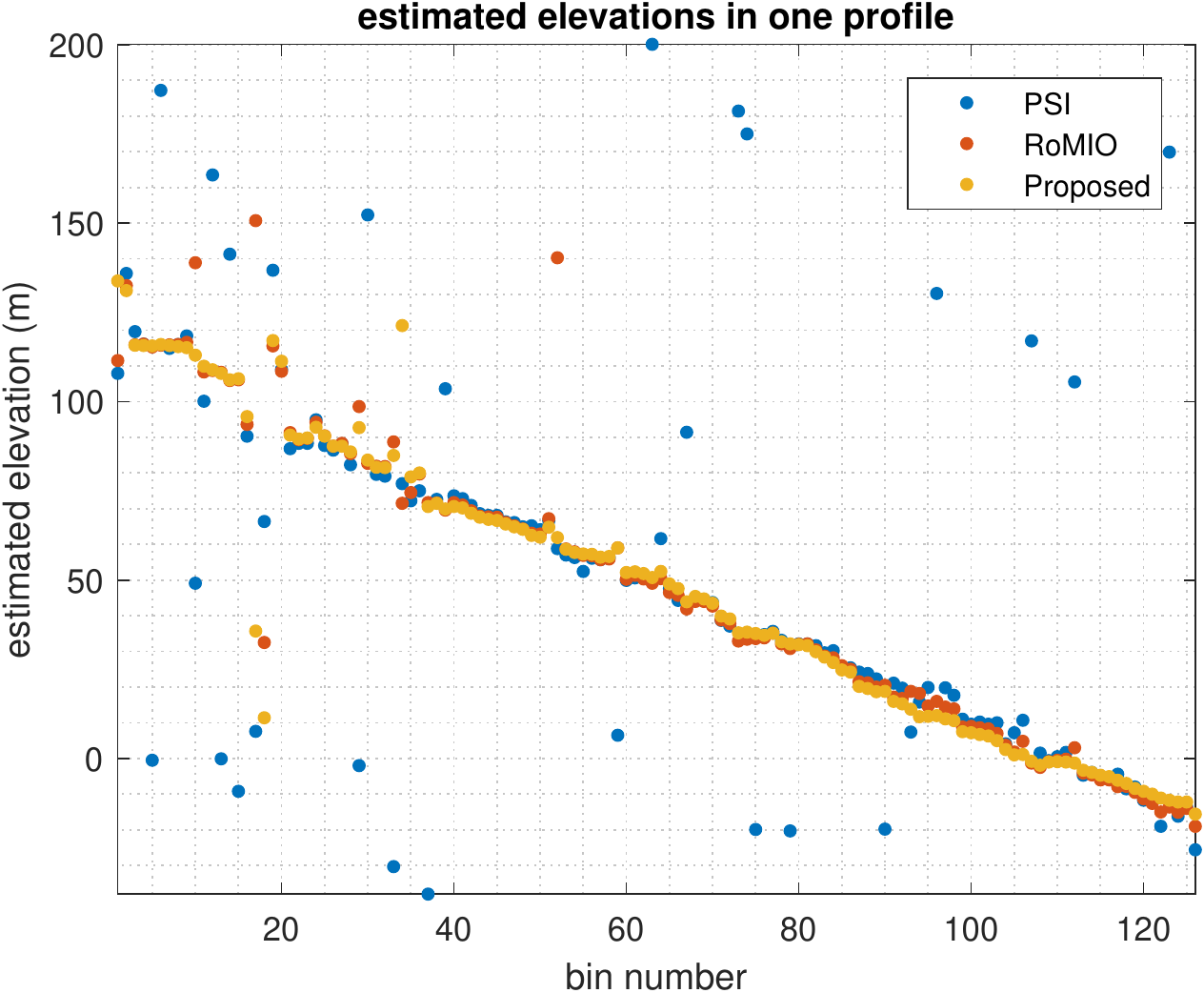}
	\caption{The extracted elevation profiles from the results shown in \Fig \ref{fg:lasvegas_ele_map} (indicated by white arrows). Besides flat areas, the geometric structure of building fa\c{c}ade can also be well preserved by the proposed method. It also gives us a hint that more accurate 3D models of urban areas can be obtained by the point cloud generated from our method.}
	\label{fg:lasvegas_ele_profile}
\end{figure}

As shown in \Fig \ref{fg:lasvegas_ele_map} and \ref{fg:lasvegas_defor_map}, consistent with the simulations, the tensor-decomposition-based methods, i.e. RoMIO and the proposed method, can achieve more robust performances than PSI. In particular, both of them can maintain reliable reconstruction results with a substack of $ 11 $ interferograms. Illustrated by the deformation estimates of Las Vegas Convention Center (\Fig \ref{fg:lasvegas_defor_map_zoomin}), many incorrectly estimated pixels of the central area on the roof exist in the PSI result. Compared to RoMIO, the proposed method can better estimate the flat roof areas. As marked by the red dashed circle, the group of noisy points is mitigated in the result of the proposed method. Moreover, the geometric structure of building fa\c{c}ade can also be well preserved by the proposed method. As illustrated in \Fig \ref{fg:lasvegas_ele_profile}, the elevation profiles are extracted from the results in \Fig \ref{fg:lasvegas_ele_map} (indicated by the white arrows). It is obvious that more noisy points exist in the result of PSI than RoMIO and the proposed method. It also gives us a hint that more accurate 3D models of urban areas can be obtained by the point cloud generated from our method. Besides, the histograms of temporal coherences are displayed in \Fig \ref{fg:temcoh_assess} (Left) based on the three reconstructed results. We can see that the fitness between the filtered InSAR data stack by our method and the model does apparently increase and there are more highly coherent points in the proposed method than RoMIO. Moreover, to further assess the reconstruction quality of the proposed method, the parameters estimated by the proposed method on the full InSAR stack were regarded as the reference, in order to compare the results of the three methods applying on a smaller InSAR stack with $ 11 $ interferorgams. The performance is demonstrated in Table \ref{tb:lasvegas_SD_bias_wrt_fullstack}. It can be seen that the proposed method can achieve more reliable estimates of geophysical parameters than both RoMIO and PSI.

\subsubsection{Berlin}
\begin{figure*}
	\centering
	\includegraphics[width=\textwidth]{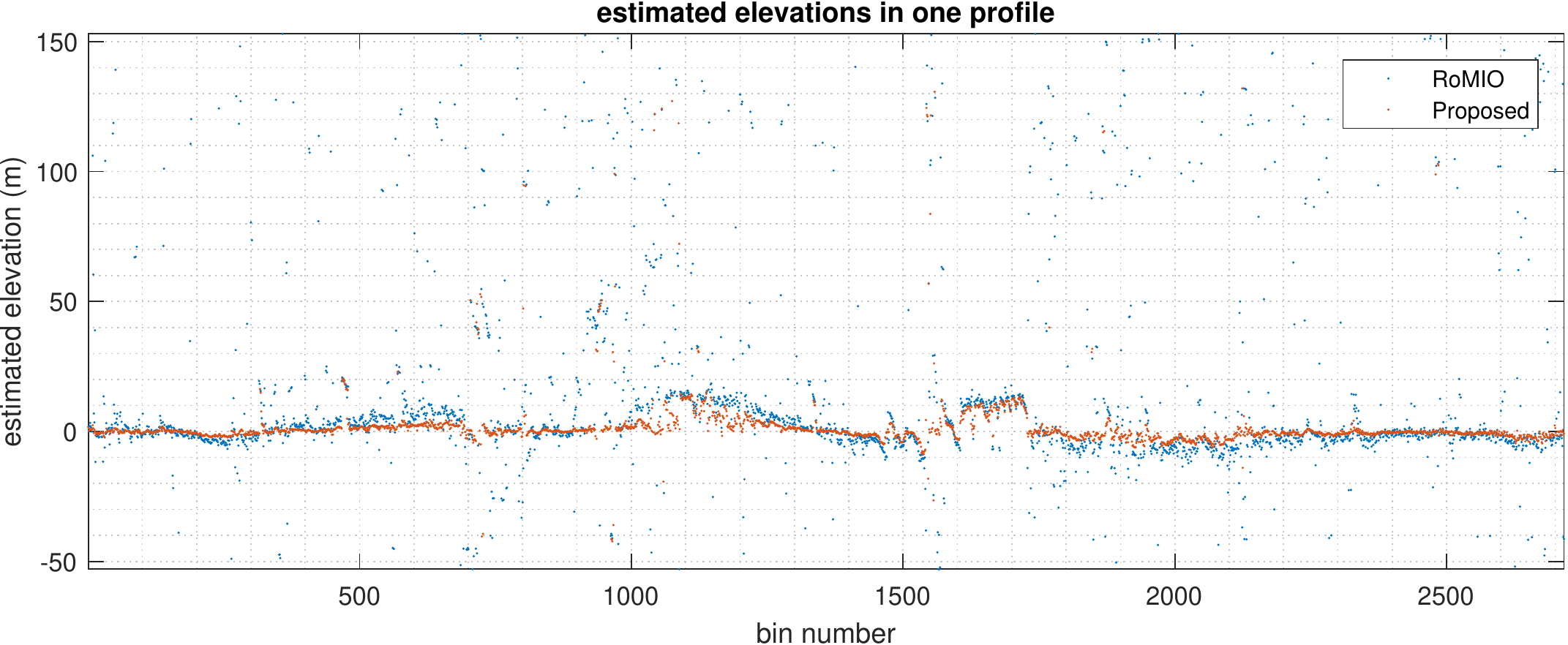}
	\caption{The extracted elevation profiles from the results shown in \Fig \ref{fg:berlin_ele_map} (indicated by red curve). Obviously, the proposed TV regularized tensor decomposition method can better preserve piecewise smoothness for the 3D reconstruction of roads than RoMIO.}
	\label{fg:berlin_road_profile}
	
\end{figure*}

\begin{figure*}
	\centering
	\includegraphics[width=0.5\textwidth]{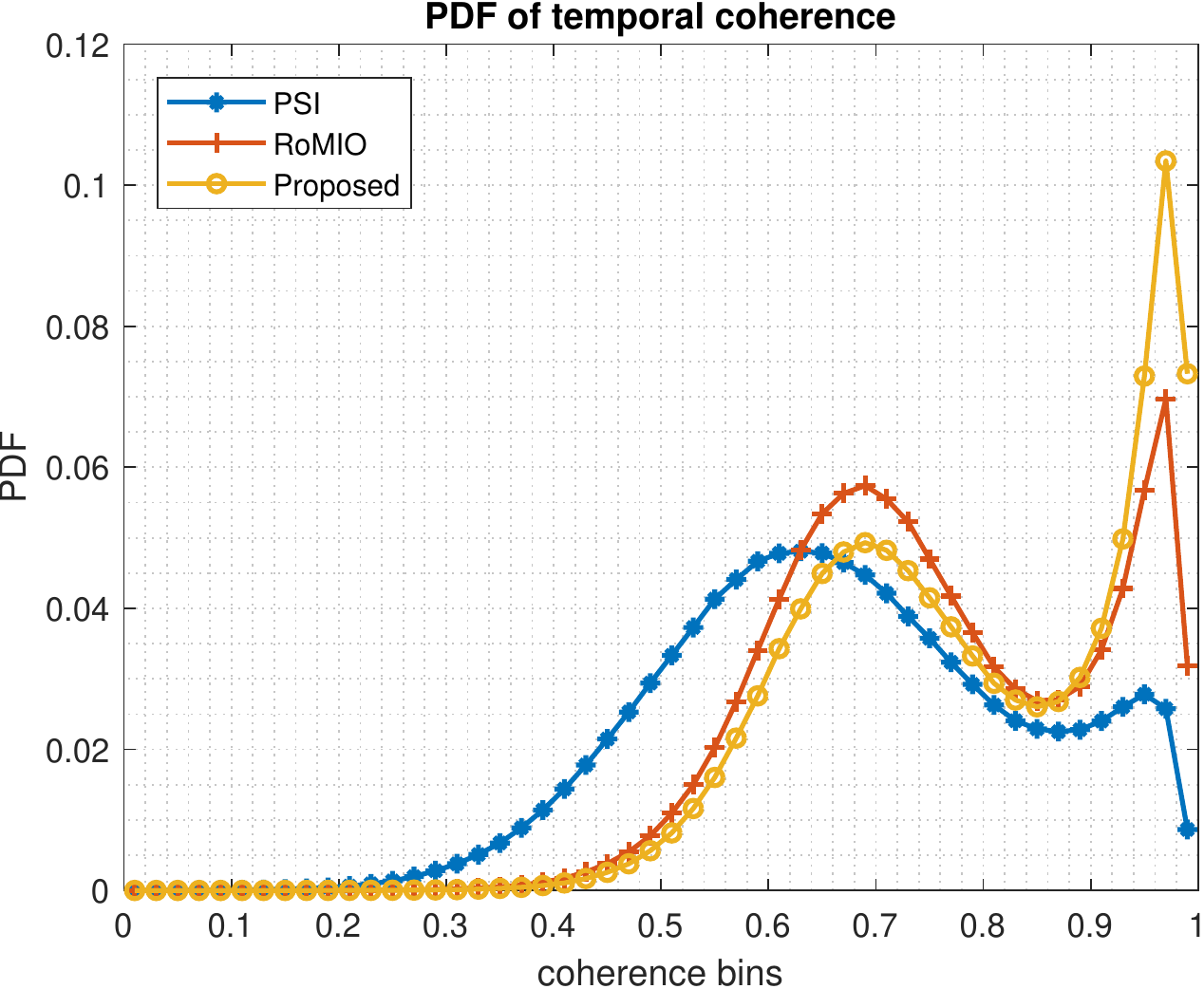}~
	\includegraphics[width=0.5\textwidth]{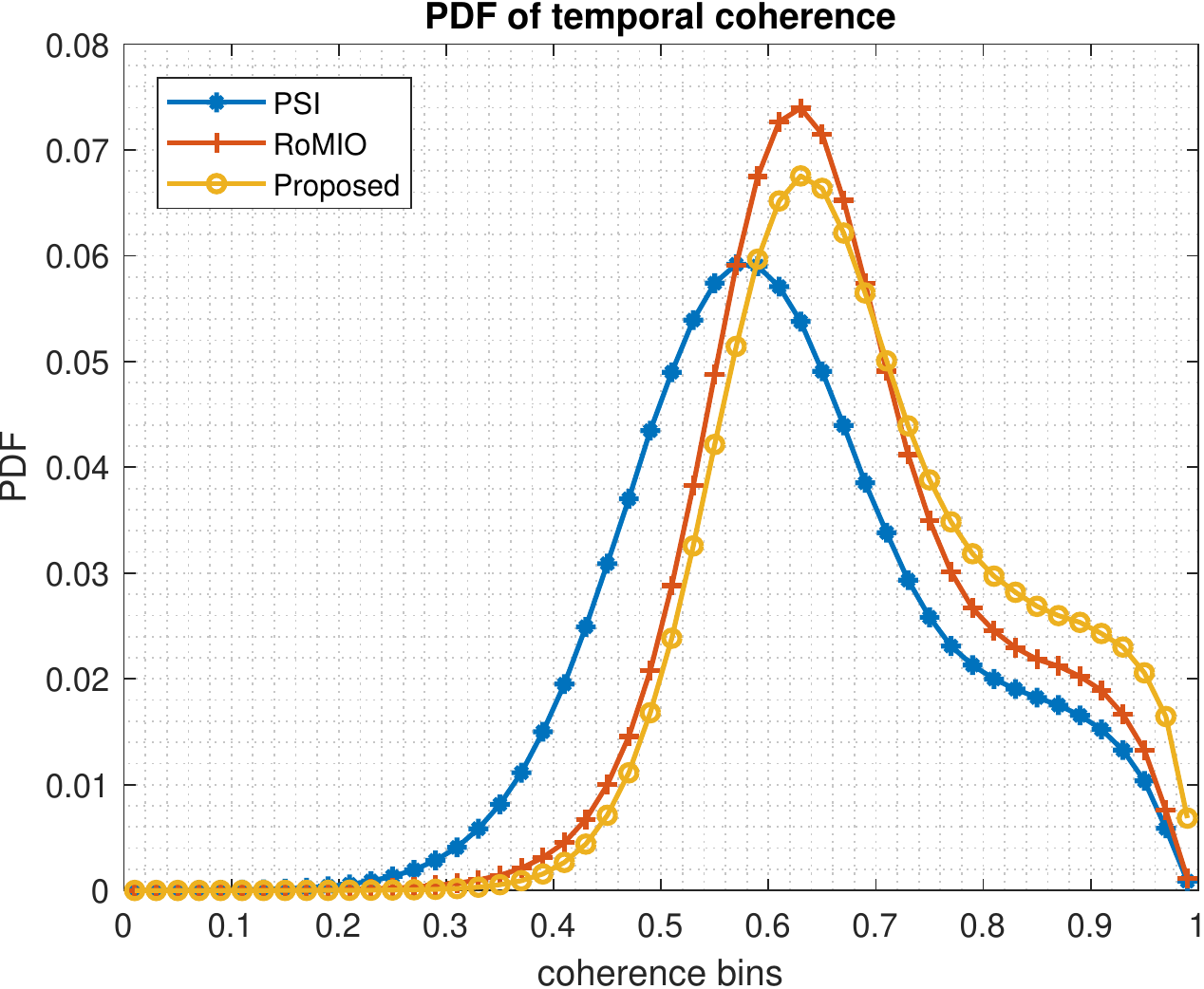}
	\caption{Probability density functions (PDF) of temporal coherences based on the estimated results by PSI and the proposed methods. (Left) shows the case of Las Vegas. (Right) shows the case of Berlin.}
	\label{fg:temcoh_assess}
\end{figure*}

From the study area shown in \Fig \ref{fg:berlin_SAR_img}, we can see it is mainly composed by building blocks and high-rise buildings. As demonstrated in \Fig \ref{fg:berlin_ele_map} and one zoom-in area in \Fig \ref{fg:berlin_ele_map_zoomin}, more outliers appear in the 3D reconstruction by PSI than RoMIO and the proposed method. Compared to RoMIO, the proposed method can better reconstruct road areas, since smoothness structure is able to be preserved by TV regularization. As an example shown in \Fig \ref{fg:berlin_ele_map} (Middle), one road profile indicated by the red curve is extracted from the results of RoMIO and the proposed method, respectively, and displayed in \Fig \ref{fg:berlin_road_profile}. Obviously, piecewise smooth property can be better maintained in the proposed method than RoMIO. Moreover, \Fig \ref{fg:berlin_defor_map} shows that the proposed method can produce the smoothest map of deformations than RoMIO and PSI, which indicates that incorrectly estimates can be mitigated by the proposed method. Consistent with the previous experiment, the filtered InSAR stack by the proposed method can best fit the model among the three comparing methods, which is displayed by the histograms of temporal coherences in \Fig \ref{fg:temcoh_assess} (Right). Besides, the numerical analysis is done in the same manner as the above experiment. As illustrated in Table \ref{tb:berlin_SD_bias_wrt_fullstack}, the estimates from the proposed method are much closer than the other two methods given the estimates from the full stack.  

\begin{table}
	\centering
	\caption{Quantitative study for the results of Las Vegas data. The parameters estimated by the proposed method on the full InSAR stack were regarded as the reference, in order to compare the results of the three methods applying on a smaller InSAR stack with $ 11 $ interferorgams.}
	\label{tb:lasvegas_SD_bias_wrt_fullstack}
	\begin{tabular}{ c| c| c| c| c}
		\hline
		\hline
		 & \multicolumn{2}{c|}{Deformation [mm/y]} & \multicolumn{2}{c}{Elevation [m]} \\
		\cline{2-5}
		 & SD & bias & SD & bias \\
		\hline
		PSI  & $ 6.06 $ & $ -1.28 $ & $ 39.56 $ & $ 11.11 $ \\
		\hline
		RoMIO & $ 1.35 $ & $ -0.51 $ & $ 7.26 $ & $ \mathbf{-0.12} $ \\
		\hline
		Proposed &  $ \mathbf{0.79} $ & $ \mathbf{-0.42} $ & $ \mathbf{2.28} $ & $ 0.44 $\\
		\hline
		\hline
	\end{tabular}
\end{table}

\begin{table}
	\centering
	\caption{Quantitative study for the results of Berlin data. The parameters estimated by the proposed method on the full InSAR stack were regarded as the reference, in order to compare the results of the three methods applying on a smaller InSAR stack with $ 15 $ interferorgams.}
	\label{tb:berlin_SD_bias_wrt_fullstack}
	\begin{tabular}{ c| c| c| c| c}
		\hline
		\hline
		& \multicolumn{2}{c|}{Deformation [mm]} & \multicolumn{2}{c}{Elevation [m]} \\
		\cline{2-5}
		& SD & bias & SD & bias \\
		\hline
		PSI  & $ 4.45 $ & $ -0.06 $ & $ 36.04 $ & $ 10.94 $ \\
		\hline
		RoMIO & $ 2.87 $ & $ \mathbf{0.02} $ & $ 23.66 $ & $ 4.30 $\\
		\hline
		Proposed &  $ \mathbf{0.80} $ & $ 0.03 $ & $ \mathbf{5.26} $ & $ \mathbf{-0.06} $\\
		\hline
		\hline
	\end{tabular}
\end{table}

\subsection{Comparison with Object-based InSAR \cite{kang2017robust}}

\begin{figure}
	\centering
	\includegraphics[width=0.45\textwidth]{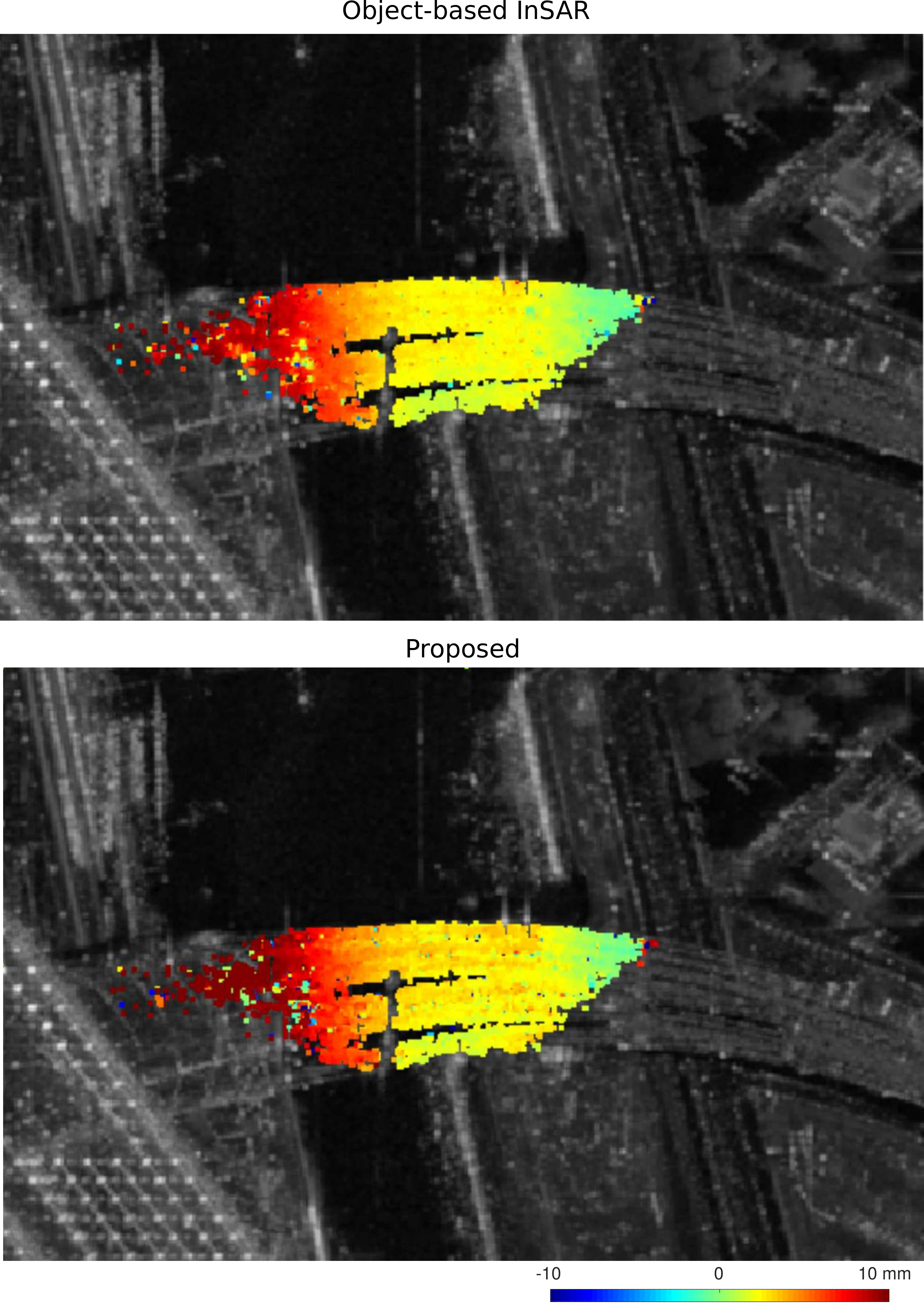}
	\caption{(Top) Amplitudes of seasonal motion estimation on one bridge in Berlin based on the method in \cite{kang2017robust}. (Down) The result based on the proposed method in this paper.}
	\label{fg:compare_with_object_based}
\end{figure}

The object-based approach in \cite{kang2017robust} contains two separate stages for the geophysical parameter estimation: tensor robust principle component analysis and the TV regularized parameter estimation. Differently with the previous approach, the proposed method integrates the two prior knowledge, i.e. the variation and low rank, into a single-stage processing. To compare the efficiencies of the two methods, we choose the same real dataset used in \cite{kang2017robust}, i.e. one bridge in the central area in Berlin. As illustrated in \Fig \ref{fg:compare_with_object_based}, it can be seen that the two methods can achieve comparable performance. Such result in turn supports our motivation of this paper: the separated optimization steps of low rank tensor decomposition and TV regularization in \cite{kang2017robust} can be merged into a single optimization. Afterwards, the estimation of height and deformation  can be done pixel by pixel, which can avoid the requirement of explicit semantic masks required in \cite{kang2017robust}. This is an advantage for code parallelization in large areas processing. 

\section{Conclusion}\label{sc:conclustion}
This paper proposed a novel tensor decomposition method in complex domain based on the prior knowledge of the low rank property and smoothness structure in multipass InSAR data stacks. Based on the proposed method, geophysical parameter estimation can be improved in real data cases, compared with conventional methods, such as PSI, and also recently proposed method --- RoMIO. Demonstrated by the case study, compared with PSI, the proposed method can improve the parameter estimation by a factor of more than seven for Berlin, and ten for Las Vegas. Furthermore, this work is the first to demonstrate tensor-decomposition-based multipass InSAR techniques can be beneficial for large-scale urban mapping problems using InSAR, including 3D urban reconstruction and surface displacement monitoring.

The proposed method introduces three parameters to be set, i.e. $ \alpha $, $ \beta $ and $ \gamma $. They do not need to be tuned simultaneously, since one parameter can be set as a constant and adjust the other two with respect to it. Based on our experiments, $ \gamma $ can be selected as a constant of $ 100/\sqrt{I_1I_2} $, the optimal $ \alpha $ lies in the range from $ 0 $ to $ 10 $, and the optimal $ \beta $ can be selected from $ 0 $ to $ 0.2 $. From the results of the real data, the proposed method is not only favorable for 3D reconstruction of flat urban areas, such as Las Vegas, but also promising for complicated European cities, such as Berlin. Moreover, for large-scale processing, the proposed method can be easily parallelized and operated in a sliding window manner. 

Since the proposed method is based on the assumption that the signals are similar both in spatial and time domain, the reconstruction for irregular signals e.g. a breakpoint or a sudden jump in deformation signals, may not be satisfied to the proposed optimization model. The results based on the proposed method favor the smoothness reconstruction and such sparse signals may be "inpainted" according to the neighboring signals in spatial and time domain.

As a future work,  we will combine the proposed method with more advanced multipass InSAR method such as D-TomoSAR, in order to produce more accurate 3D reconstruction in urban areas. The improved 3D reconstruction can be a great input to the urban 3D model reconstruction \cite{shahzad2015robust}. Moreover, since atmosphere phase screens (APS) in multi-temporal InSAR stack partly fulfill the assumption of the proposed model (i.e. APS is only spatially correlated but not temporally), it would be interesting to systematically investigate the performance of atmosphere signal removal based on such tensor-decomposition based method.

% or


\appendices
\section{ADMM solver for \eqref{eq:3DTVLR}}
%% you can choose not to have a title for an appendix
%% if you want by leaving the argument blank
%\section{}
\begin{algorithm}
	\caption{Problem \eqref{eq:3DTVLR} solved by ADMM}
	\begin{algorithmic}[1]
		\Require $ \mathcal{G},\alpha,\beta,\gamma,N $ % input 
		\State Initialize $ \mathcal{X}=\mathcal{E}=\mathcal{Z}=\mathcal{F}=\mathcal{T}_1=\mathcal{T}_2=\mathcal{T}_3=0 $, $ \mu_{\max}=10^{10} $, $ \eta=1.1 $, $ \mu=10^{-2} $
		\For {$ k=0 $ to $ \mathrm{maxIter} $}
		\State Update $ \mathcal{X}^{(k+1)} $ by SVT for mode-$ n $ unfolding matrix of $ \frac{1}{2}(\mathcal{G}-\mathcal{E}^{(k)}+\mathcal{Z}^{(k)}+\frac{\mathcal{T}_1^{(k)}-\mathcal{T}_2^{(k)}}{\mu}) $, \par 
		then mode-$ n $ folding of the results as $ N $ tensors and average them by $ N $:\par
		$ \mathcal{X}^{(k+1)}\leftarrow\frac{1}{N}\sum_{n=1}^{N}\mathcal{S}_{n,\beta{N}/\mu}(\frac{1}{2}(\mathbf{G}_{(n)}-\mathbf{E}_{(n)}^{(k)}+\mathbf{Z}_{(n)}^{(k)}+\frac{\mathbf{T}_{1(n)}^{(k)}-\mathbf{T}_{2(n)}^{(k)}}{\mu})) $,
		where $ \mathcal{S}_{n,\beta{N}/\mu}(\cdot):=\mathrm{fold}_n(\mathcal{S}_{\beta{N}/\mu}(\cdot)) $.
		\State Update $ \mathcal{Z}^{(k+1)} $ by calculating $ H_{\mathcal{Z}} $ and  $ T_{\mathcal{Z}} $, where \par 
		$ H_{\mathcal{Z}}=\mathcal{T}_2^{(k)}-D^*(\mathcal{T}_3^{(k)})+\mu\mathcal{X}^{(k+1)}+\mu{D^*}(\mathcal{F}^{(k)}) $ and \par 
		$ T_{\mathcal{Z}}=|\mathrm{fftn}(D_1)|^2+|\mathrm{fftn}(D_2)|^2+|\mathrm{fftn}(D_3)|^2 $, \par
		$ \mathcal{Z}^{(k+1)}\leftarrow\mathrm{ifftn}(\frac{\mathrm{fftn}(H_{\mathcal{Z}})}{\mu\mathbf{I}+\mu T_{\mathcal{Z}}}) $.
		\State Update $ \mathcal{F}^{(k+1)} $ by element-wise soft-thresholding of tensor $ D(\mathcal{Z}^{(k+1)})+\mathcal{T}_3^{(k)}/\mu $: \par
		$ \mathcal{F}^{(k+1)}\leftarrow\mathcal{R}_{\alpha/\mu}(D(\mathcal{Z}^{(k+1)})+\mathcal{T}_3^{(k)}/\mu) $.
		
		\State Update $ \mathcal{E}^{(k+1)} $ by element-wise soft-thresholding of tensor $ \mathcal{G}+\mathcal{T}_1^{(k)}/\mu-\mathcal{X}^{(k+1)} $: \par
		$ \mathcal{E}^{(k+1)}\leftarrow\mathcal{R}_{\gamma/\mu}(\mathcal{G}+\mathcal{T}_1^{(k)}/\mu-\mathcal{X}^{(k+1)}) $.
		\State Update $ \mathcal{T}_1^{(k+1)} $, $ \mathcal{T}_2^{(k+1)} $ and $ \mathcal{T}_3^{(k+1)} $ by  \par 
		$ \mathcal{T}_1^{(k+1)}\leftarrow\mathcal{T}_1^{(k)}+\mu(\mathcal{G}-\mathcal{X}^{(k+1)}-\mathcal{E}^{(k+1)}) $, \par
		$ \mathcal{T}_2^{(k+1)}\leftarrow\mathcal{T}_2^{(k)}+\mu(\mathcal{X}^{(k+1)}-\mathcal{Z}^{(k+1)}) $, \par 
		$ \mathcal{T}_3^{(k+1)}\leftarrow\mathcal{T}_3^{(k)}+\mu(D(\mathcal{Z}^{(k+1)})-\mathcal{F}^{(k+1)})$.
		\State Update $ \mu $ by $ \mu\leftarrow\min(\eta\mu,\mu_{\max}) $.
		\If{convergence}
		\State	break
		\EndIf
		\EndFor
		\Ensure $ (\hat{\mathcal{X}},\hat{\mathcal{E}}) $% output
	\end{algorithmic}
	\label{ag:ADMM_solver}
\end{algorithm}

%\section*{Acknowledgment}

\ifCLASSOPTIONcaptionsoff
  \newpage
\fi



\bibliographystyle{IEEEtran}
\bibliography{bibliography}

\begin{IEEEbiography}[{\includegraphics[width=1in,height=1.25in,clip,keepaspectratio]{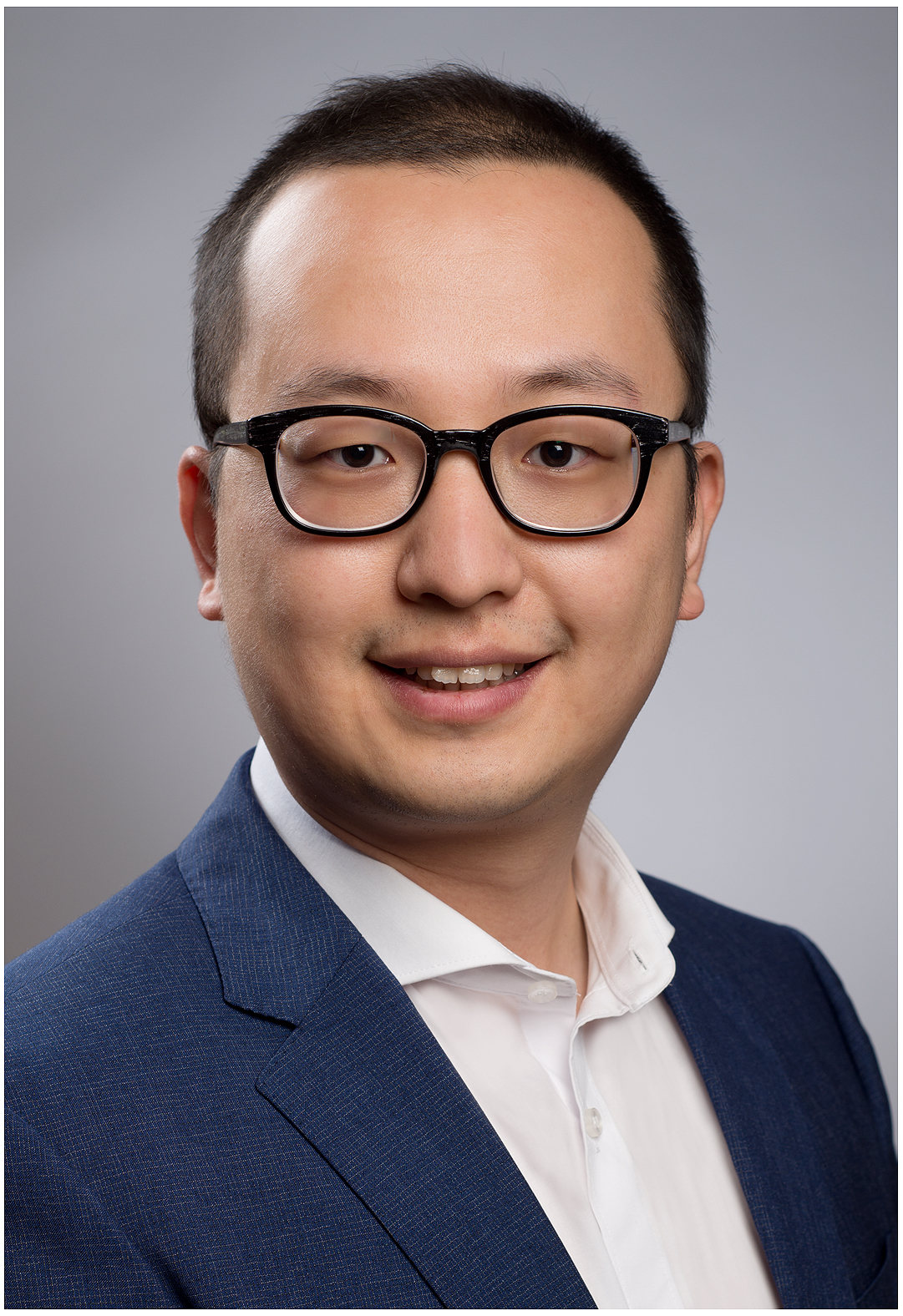}}]{Jian Kang}
	(S'16-M'19) received B.S. and M.E. degrees in electronic engineering from Harbin Institute of Technology (HIT), Harbin, China, in 2013 and 2015, respectively, and Dr.-Ing. degree from Signal Processing in Earth Observation (SiPEO), Technical University of Munich (TUM), Munich, Germany, in 2019. In August of 2018, he was a guest researcher at Institute of Computer Graphics and Vision (ICG), TU Graz, Graz, Austria. He is currently with Faculty of Electrical Engineering and Computer Science, Technische Universit\"at Berlin (TU-Berlin), Berlin, Germany. His research focuses on signal processing and machine learning, and their applications in remote sensing. In particular, he is interested in multi-dimensional data analysis, geophysical parameter estimation based on InSAR data, SAR denoising and deep learning based techniques for remote sensing image analysis.
	
	He obtained first place of the best student paper award in EUSAR 2018, Aachen, Germany.
\end{IEEEbiography}

\begin{IEEEbiography}[{\includegraphics[width=1in,height=1.25in,clip,keepaspectratio]{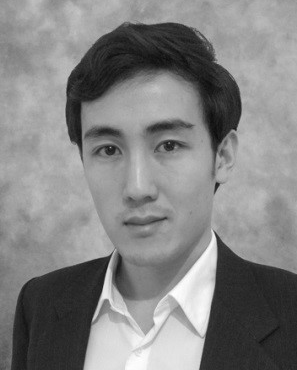}}]{Yuanyuan Wang}
	(S'10-M'14) received his B.Eng. (Hons.) degree in electrical engineering from The Hong Kong Polytechnic University, Hong Kong, China, in 2008, and the M.Sc. degree as well as the Dr.-Ing. degree from Technical University of Munich (TUM), Munich, Germany in 2010 and 2015, respectively. In June and July of 2014, he was a guest scientist at the Institute of Visual Computing, ETH Zurich, Switzerland. Currently, he is with Signal Processing in Earth Observation (http://www.sipeo.bgu.tum.de/), TUM, as well as with the Department of EO Data Science of German Aerospace Center, where he leads the working group ``Big SAR Data''. His research interests include optimal and robust parameters estimation in multibaseline InSAR techniques, multisensor fusion algorithms of SAR and optical data, non-linear optimization with complex numbers, machine learning in SAR, and high performance computing for big data. He was one of the best reviewers of IEEE TGRS 2016.
\end{IEEEbiography}

\begin{IEEEbiography}[{\includegraphics[width=1in,height=1.25in,clip,keepaspectratio]{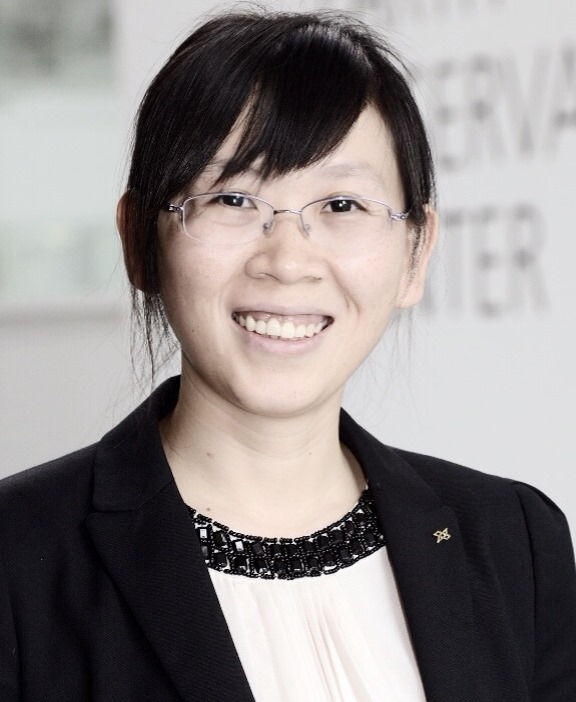}}]{Xiao Xiang Zhu}(S'10--M'12--SM'14) received the Master (M.Sc.) degree, her doctor of engineering (Dr.-Ing.) degree and her ``Habilitation'' in the field of signal processing from Technical University of Munich (TUM), Munich, Germany, in 2008, 2011 and 2013, respectively.
\par
  She is currently the Professor for Signal Processing in Earth Observation (www.sipeo.bgu.tum.de) at Technical University of Munich (TUM) and German Aerospace Center (DLR); the head of the department ``EO Data Science'' at DLR's Earth Observation Center; and the head of the Helmholtz Young Investigator Group ``SiPEO'' at DLR and TUM. Since 2019, Zhu is co-coordinating the Munich Data Science Research School (www.mu-ds.de). She is also leading the Helmholtz Artificial Intelligence (www.haicu.de) -- Research Field ``Aeronautics, Space and Transport". Prof. Zhu was a guest scientist or visiting professor at the Italian National Research Council (CNR-IREA), Naples, Italy, Fudan University, Shanghai, China, the University  of Tokyo, Tokyo, Japan and University of California, Los Angeles, United States in 2009, 2014, 2015 and 2016, respectively. Her main research interests are
  remote sensing and Earth observation, signal processing, machine learning and data science, with a special application focus on global urban mapping.

  Dr. Zhu is a member of young academy (Junge Akademie/Junges Kolleg) at the Berlin-Brandenburg Academy of Sciences and Humanities and the German National  Academy of Sciences Leopoldina and the Bavarian Academy of Sciences and Humanities. She is an associate Editor of IEEE Transactions on Geoscience and Remote Sensing.
\end{IEEEbiography}

\vfill


\end{document}